\documentclass[useAMS,usenatbib]{mn2e}
\usepackage{graphicx, txfonts, natbib}
\usepackage{epsf}
\usepackage{amssymb}
\usepackage{epstopdf}
\usepackage{longtable}
\usepackage{subfigure}
\usepackage{comment}
\usepackage[colorlinks=false,dvips]{hyperref}
\newcommand{\msun}{\mbox{M$_{\odot}$}}

\newcommand{\kms}{\mbox{$\rm{km}\,s^{-1}$}}

\DeclareMathAlphabet{\mathsc}{OT1}{cmr}{m}{sc}
\def\testbx{bx}%
\DeclareRobustCommand{\ion}[2]{%
\relax\ifmmode
\ifx\testbx\f@series
{\mathbf{#1\,\mathsc{#2}}}\else
{\mathrm{#1\,\mathsc{#2}}}\fi
\else\textup{#1\,{\mdseries\textsc{#2}}}%
\fi}

\newcommand{\Nai}{\ion{Na}{i}}

\newcommand{\Caii} {\ion{Ca}{ii}}

\newcommand{\SiII} {\ion{Si}{ii}}

\newcommand{\Ki}{\ion{K}{i}}

\title[CSM studies of SNe Ia] {A statistical analysis of circumstellar material in Type Ia supernovae}
 \author[K. Maguire et al.]
 {K.~Maguire,$^1$\thanks{E-mail: kate.maguire@astro.ox.ac.uk} M.~Sullivan,$^{2}$  F.~Patat,$^3$ A.~Gal-Yam,$^4$  I.~M.~Hook,$^{1,5}$ S.~Dhawan,$^{1}$  \newauthor D.~A.~Howell,$^{6,7}$
 P.~Mazzali,$^{8}$  P.~E.~Nugent,$^{9,10}$   Y.-C.~Pan,$^1$ P.~Podsiadlowski,$^1$  J.~D.~Simon,$^{11}$  \newauthor A.~Sternberg,$^{12}$ S.~Valenti,$^{6,7}$ C.~Baltay,$^{13}$
D.~Bersier,$^{8}$  N.~Blagorodnova,$^{14}$   T.-W.~Chen,$^{15}$   \newauthor N.~Ellman,$^{13}$ U.~Feindt,$^{16}$  F.~F\"orster,$^{17}$  M.~Fraser,$^{15}$ 	S.~Gonz\'alez-Gait\'an,$^{17}$   M.~L.~Graham,$^{6,7}$    \newauthor    C.~Guti\'errez,$^{17}$ S.~Hachinger,$^{18}$   E.~Hadjiyska,$^{13}$ C.~Inserra,$^{15}$  C.~Knapic,$^{19}$  R.~R.~Laher,$^{20}$   \newauthor G.~Leloudas,$^{21,22}$       S.~Margheim,$^{23}$ R.~McKinnon,$^{13}$  M.~Molinaro,$^{19}$ N.~Morrell,$^{24}$ E.~O.~Ofek,$^4$ \newauthor   D.~Rabinowitz,$^{13}$    A.~Rest,$^{25}$ D.~Sand,$^{26}$   R.~Smareglia,$^{19}$ S.~J.~Smartt,$^{15}$     F.~Taddia,$^{27}$ \newauthor  E.~S.~Walker,$^{13}$   N.~A.~Walton,$^{14}$ D.~R.~Young$^{15}$\\
          $^1$Department of Physics (Astrophysics), University of Oxford, DWB, Keble Road, Oxford OX1 3RH, UK\\
     $^2$Physics \& Astronomy, University of Southampton, Southampton, Hampshire, SO17 1BJ, UK\\
      $^3$European Organisation for Astronomical Research in the Southern Hemisphere (ESO), Karl-Schwarzschild-Str. 2, 85748 Garching b. M\"unchen, Germany \\
      $^4$Benoziyo Center for Astrophysics, Weizmann Institute of Science, 76100 Rehovot, Israel\\
      $^5$INAF Ð Osservatorio Astronomico di Roma, via Frascati, 33, 00040 Monte Porzio Catone, Roma, Italy\\
     $^6$Las Cumbres Observatory Global Telescope Network, Goleta, CA 93117, USA \\
     $^7$Department of Physics, University of California, Santa Barbara, CA 93106-9530, USA \\
     $^{8}$Liverpool John Moores, Liverpool Science Park, IC2 building, 146 Brownlow Hill, Liverpool L3 5RF, UK\\
     $^9$Cahill Center for Astrophysics, California Institute of Technology, East California Boulevard, Pasadena, CA 91125, USA \\
   $^{10}$Computational Cosmology Center, Lawrence Berkeley National Laboratory, 1 Cyclotron Road, Berkeley, CA 94720, USA \\
$^{11}$Observatories of the Carnegie Institution of Washington, 813 Santa Barbara St., Pasadena, CA 91101, USA\\
$^{12}$Max-Planck-Institut f\"ur Astrophysik, Karl-Schwarzschildstr. 1, D-85748 Garching, Germany \\
$^{13}$Department of Physics, Yale University, New Haven, CT 06250-8121, USA\\
     $^{14}$Institute of Astronomy, University of Cambridge, Madingley Road, Cambridge CB3 0HA, UK \\
     $^{15}$Astrophysics Research Centre, School of Mathematics and Physics, Queen's University Belfast,  Belfast, BT7 1NN, UK \\
$^{16}$Physikalisches Institut, Universit\"at Bonn, Nu$\ss$allee 12, 53115 Bonn, Germany\\
    $^{17}$Departamento de Astronom\'ia, Universidad de Chile, Casilla 36-D, Santiago, Chile\\   
    $^{18}$Julius-Maximilians-Universit\"at W\"urzburg, Emil-Fischer-Str.~31, 97074 W\"urzburg, Germany\\
$^{19}$INAF - Osservatorio Astronomico di Trieste, Via G.B. Tiepolo 11, I-34143 Trieste, Italy\\
$^{20}$Spitzer Science Center, California Institute of Technology,  M/S 314-6, Pasadena, CA 91125, U.S.A.\\
$^{21}$The Oskar Klein Centre, Department of Physics, Stockholm University, AlbaNova, 10691 Stockholm, Sweden \\
$^{22}$Dark Cosmology Centre, Niels Bohr Institute, University of Copenhagen, Juliane Maries Vej 30, 2100 Copenhagen, Denmark \\
$^{23}$Gemini Observatory, Southern Operations Center, Casilla 603, La Serena, Chile\\
$^{24}$Las Campanas Observatory, Carnegie Observatories, Casilla 601, La Serena, Chile\\
$^{25}$Space Telescope Science Institute, 3700 San Martin Drive, Baltimore, MD 21218, USA\\
$^{26}$Texas Tech University, Physics Department, Box 41051, Lubbock, TX 79409-1051, USA\\
$^{27}$The Oskar Klein Centre, Department of Astronomy, Stockholm University, AlbaNova, 10691 Stockholm, Sweden\\}

\begin{document} 
\maketitle
\clearpage
\begin{abstract}
A key tracer of the elusive progenitor systems of Type Ia supernovae
  (SNe Ia) is the detection of narrow blueshifted time-varying \Nai\ D
  absorption lines, interpreted as
  evidence of circumstellar material (CSM) surrounding the progenitor system.
  The origin of this material is controversial, but the simplest
  explanation is that it results from previous mass loss in a system containing a white
  dwarf and a non-degenerate companion star.  We present new single-epoch
  intermediate-resolution spectra of 17 low-redshift SNe Ia taken with
 XShooter on the ESO Very Large Telescope.   Combining this sample
  with events from the literature, we confirm an excess ($\sim$20 per cent) of SNe Ia displaying blueshifted narrow \Nai\ D absorption features compared to non-blueshifted \Nai\ D features. The host galaxies of SNe Ia displaying blueshifted absorption profiles are skewed towards later-type galaxies, compared to SNe Ia that show no \Nai\ D absorption, and SNe Ia displaying blueshifted narrow \Nai\ D absorption features have broader light curves.
 The strength of the \Nai\ D absorption is stronger in SNe Ia  displaying blueshifted  \Nai\ D absorption features than those without blueshifted features, and the strength of the blueshifted \Nai\ D is correlated with the $B-V$ colour of the SN at maximum light. This strongly suggests the absorbing material is local to the SN. In the context of the progenitor systems of SNe Ia, we discuss the significance of these findings and other recent observational evidence on the nature of SN Ia progenitors. We present a summary that suggests there are at least two distinct populations of Ônormal, cosmologically usefulÕ SNe Ia.
 \end{abstract}
\begin{keywords}
distance scale -- supernovae: general -- circumstellar matter
\end{keywords}

\section{Introduction} \label{intro}

Type Ia supernovae (SNe Ia) are excellent standardisable candles and
play an important role in constraining cosmological parameters
\citep[e.g.,][]{1998AJ....116.1009R,2007ApJ...659...98R,1999ApJ...517..565P,2009ApJS..185...32K,2011ApJ...737..102S,2012ApJ...746...85S}.
However, there is still much debate over the nature of their progenitor
systems. SNe Ia have long been suspected to be the result of the
thermonuclear explosion of an accreting carbon-oxygen white dwarf (CO-WD) 
in a close binary system, and the compact nature of the exploding star
has recently been confirmed observationally
\citep{2011Natur.480..344N,2012ApJ...744L..17B}. However, the nature of the
companion to the white dwarf remains unknown. Two main types of
progenitor systems are generally considered: the double degenerate
\citep[DD;][]{1984ApJS...54..335I,1984ApJ...277..355W} scenario, with
two white dwarfs, and the single degenerate
\citep[SD;][]{1973ApJ...186.1007W} scenario, with a non-degenerate
companion star to the white dwarf, such as a giant, sub-giant or
main-sequence star. Recent
observational evidence suggests that both of these channels may
operate
\citep{2010Natur.463..924G,2011Natur.480..348L,2011Sci...333..856S,2012Sci...337..942D,2012Natur.481..164S},
although the relative frequency of each and the relation to host galaxy environment is not yet clear.

One observational tracer of different progenitor configurations is the
detection of narrow time-varying blueshifted \Nai\ D absorption features in SN Ia spectra \citep{2007Sci...317..924P,2009ApJ...702.1157S,2009ApJ...693..207B,2010AJ....140.2036S}.
Such time-varying features, reminiscent of the spectra of both classical \citep{2008ApJ...685..451W} and recurrent
novae \citep{2011A&A...530A..63P}, are suggestive of outflowing
material from the SN system, naturally explained by the SD scenario:
non-accreted material from the donor star is blown away from the
system prior to explosion and remains afterward as circumstellar material (CSM).  
The time-varying nature of the \Nai\ D profiles suggests that the material is very close to the SN ($\sim$10$^{16}$--10$^{17}$ cm) and must be associated with the progenitor system. An interstellar origin for the variations caused  by transverse proper motion in the absorbing material and/or line-of-sight effects can be ruled out since the \Caii\ H\&K absorption features would also be expected to vary, which is not observed.

It was originally assumed that in the merger of two WD, CSM material would not be present. However, recent theoretical studies looking at both violent WD-WD mergers soon after the common envelope phase \citep[][]{2013MNRAS.429.1425R,2013MNRAS.431.1541S}, as well as the interaction of ejected material from WD-WD binaries with the interstellar medium (ISM) have suggested that the DD channel may be capable of producing CSM that could be detectable in some SNe Ia \citep{2013arXiv1302.2916S,2013arXiv1304.4957R}.

\citet{2011Sci...333..856S} extended the single-object studies investigating time-varying \Nai\ D absorption features, by searching for \Nai\ D lines in single-epoch high-resolution spectra in a sample of 35
nearby SNe Ia.  They showed that SNe Ia in spiral galaxies have a
strong statistical preference for displaying blueshifted absorption structures
in their spectra.  Since only single-epoch spectra were obtained, the presence of blueshifted \Nai\ D lines in any individual event
does not directly imply the presence of CSM -- some of these blueshifted (and all of the redshifted) features
are expected to be due to the presence of interstellar absorption features.
However, there is no reason to expect a preference for blueshifted
over redshifted lines due to host galaxy material, and
\citet{2011Sci...333..856S} interpret their results as evidence for outflowing material in some SNe Ia in spiral galaxies. Using additional data from the literature combined with some new observations,
\citet{2012ApJ...752..101F} suggested that SNe Ia with blueshifted
absorption profiles have, on average, higher \SiII\ 6355 \AA\
velocities and redder colours at maximum light relative to the rest of the SN Ia population. While the redder colours could be explained by an additional contribution of CSM to the SN colour \citep[e.g.,][]{2011ApJ...735...20A}, the \SiII\ velocity trends are more difficult to explain. 

In this paper, we present the results of an observational campaign to
further link CSM signatures in SNe Ia with their photometric, spectral
and host galaxy properties. We use intermediate-resolution spectra obtained over a
multi-period programme with the European Southern Observatory (ESO) Very Large Telescope (VLT) and the XShooter
spectrograph to search for CSM signatures in a new sample of nearby
SNe Ia. These data are complemented by light curves and spectra obtained through monitoring campaigns at multiple facilities to determine the relationship between SN Ia progenitor configurations and observed SN properties. Throughout this paper we assume a Hubble constant of $H_0=70$\,km\,s$^{-1}$\,Mpc$^{-1}$.

\begin{table*}
  \caption{Discovery details and host galaxy properties of the XShooter SN Ia sample.}
 \label{tab:disc}
\begin{tabular}{@{}lccccccccccccccccccccccccccccccc}
  \hline
  \hline
 SN name&R.A. & Dec.&Galaxy name&Heliocentric&Galaxy&Disc.&Disc.&Spectroscopically\\
 	& (J2000)	&(J2000)	&			&redshift ($z_{heli}$)&	 type$^a$		& source$^b$&epoch&classified$^c$\\
\hline
\hline
LSQ12dbr	&20:58:51.89 &	-02:58:27.1&Anon.&0.0196$\pm$0.000344$^e$&Irr$^1$&LSQ&20120616 &Gemini, ATel 4212\\
LSQ12fuk&04:58:15.89 	&-16:17:57.8 &Anon.&0.02020$\pm$0.00006$^e$&Sab$^1$&LSQ&20121031& NSF-II, ATEL 4537\\
LSQ12fxd&05:22:16.99 &	-25:35:47.0&ESO 487-G 004&0.031242$\pm$0.000017$^f$&Sc&LSQ&20121101&PESSTO, ATEL 4545\\
LSQ12gdj	&23:54:43.32 &	-25:40:34.0&ESO 472- G 007&0.030324$\pm$0.000057$^f$&Sc&LSQ &20121107&NSF-II, ATEL 4566 \\
LSQ12hzj&09:59:12.43& 	-09:00:08.3&2MASX J09591230-0900095&0.029$\pm$0.001$^e$&S0$^1$&LSQ&20121224&NSF-II, ATEL 4701\\
PTF12iiq &	02:50:07.76 &-00:15:54.4	&2MASX J02500784-0016014&0.02908$\pm$0.00001$^f$&S0$^1$&PTF&20120829&PTF, ATel 4363\\
PTF12jgb&04:15:01.44& -15:20:53.7&2MASXi J0415016-152053&	0.02811$\pm$0.00008$^e$&--$^2$&PTF&20121002 &PTF, Gemini$^i$\\
SN\,2012cg$^d$&12:27:12.83&+09:25:13.1&NGC 4424&0.001538$\pm$0.000013$^g$&Sa&LOSS& 20120517&Lick, ATEL 4115\\
SN\,2012et$^d$ & 23:42:38.82&+27 05 31.5&CGCG 476-117& 0.02478$\pm$0.00002$^e$&Sb&TOCP&20120912&Asiago, CBET 3227\\
SN\,2012fw$^d$&21:01:58.99&-48:16:25.9&ESO 235-37&0.018586$\pm$0.000150$^f$&S0/a&TAROT&20120819&PESSTO, ATEL 4339\\
SN\,2012hd$^d$ & 01:14:07.46 &-32:39:07.7&IC 1657&0.01218$\pm$0.00002$^e$& Sbc&TOCP&20121121&PESSTO, ATEL 4602\\
SN\,2012hr$^d$&06:21:38.46&-59:42:50.6& ESO 121-26&0.007562$\pm$0.000013$^f$&Sbc&TOCP&20121217&CSP, ATEL 4663\\
SN\,2012ht$^d$&10:53:22.75&+16:46:34.9&	NGC 3447&0.003559$\pm$0.00004$^f$&Irr$^1$&TOCP&20121219&Asiago, CBET 3350\\
SN\,2013U$^d$&10:01:12.00&+00:19:42.3&CGCG 8-23 &0.03417$\pm$0.00008$^e$&Sc&TOCP&20130205&Asiago, ATEL 4796\\
SN\,2013aa$^d$&14:32:33.88&-44:13:27.8&NGC 5643 &0.003999$\pm$0.000007$^f$&Sc&TOCP&20130214&FLOYDS, ATEL 4817\\
SN\,2013aj$^d$&13:54:00.68&-07:55:43.8&NGC 5339&0.009126$\pm$0.000010$^f$&Sa pec&TOCP&20130303&PESSTO, ATEL 4852 \\
SN\,2013ao$^d$&11:44:44.74&-20:31:41.1&Anon.&$\sim$0.04$^h$&Dwarf$^1$&CRTS&20130304&PESSTO, ATEL 4863\\
 \hline
\end{tabular}
 \begin{flushleft}
$^a$Source of galaxy classification is NED unless otherwise noted. $^1$Visually classified for this paper. $^2$Classification of the host of PTF12jgb is unclear from available images. \\
$^b$Discovery source description is given in Section \ref{sec:sample-selection}\\
$^c$The telescope or collaboration that spectroscopically classified the SN. Further details can be found in the listed ATEL or CBET. \\
$^d$Alternative names and additional references: SN\,2012cg \citep{2012ApJ...756L...7S}, SN\,2012et = PSN J23423882+2705315 (CBET 3227), SN\,2012fw = PSN J21015899-4816259 (CBET 3282), SN\,2012hd = LSQ12gqn (CBET 3324), SN\,2012hr = PSN J06213846-5942506 (CBET 3346), SN\,2012ht = PSN J10532275+1646349, SN\,2013U =PSN J10011200+0019423 (CBET 3410), SN\,2013aa = PSN J14323388-4413278 (CBET 3416), SN\,2013aj = PSN J13540068-0755438 (CBET 3434),  SN\,2013ao = SSS130304:114445-203141 (ATEL 4908, CBET 3442).  \\
 $^e$Redshift measured from host galaxy features in SN spectrum.
 $^f$Redshift from recessional velocity obtained from NED or SDSS Data Release 9.\\
 $^g$Redshift calculated from the stellar velocity field of \citet{2006AJ....131..747C} at the position of SN\,2012cg \\
 $^h$A redshift for SN\,2013ao is not listed in NED/SDSS and no host galaxy lines are identified in any of its spectra. A redshift of $\sim$0.04 is obtained from SN spectral fitting but does not have the necessary accuracy for this study. \\
 $^i$No ATEL or CBET was released for PTF12jgb. It was classified at Gemini North using the Gemini Multi-Object Spectrograph (GMOS) by the PTF collaboration on 20121004 as a SN Ia.\\
 \end{flushleft}
\end{table*}

\section{Observations and Data Reduction}
\label{obs_data}
 
In this section, we present new observations of 17 SNe Ia that include intermediate-resolution spectra, complemented by low-resolution spectra and light
curve data. We combine our new data set with 16 events from the
literature, giving a total sample of 33 events. We discuss the
sample selection, spectroscopic observations, and photometric
monitoring in turn.

\subsection{Sample selection}
\label{sec:sample-selection}

The new SN Ia data were obtained over the course of a multi-period
programme at the VLT using the XShooter spectrograph
\citep{2011A&A...536A.105V}.  Details of the 17 SNe Ia, discovered by a variety of surveys/searches, are listed in
Table~\ref{tab:disc}. The SNe were selected according to the following two criteria: i)
that the SNe were located at $z<0.03$ to enable a high signal-to-noise
spectrum to be obtained, and ii) that the SNe were spectroscopically classified as SNe Ia prior to
maximum light, so that a reliable light curve could be measured. We did not preferentially select SNe Ia displaying redder optical colours or strong \Nai\ D absorption features in the low resolution classification spectra (as was historically the case) to ensure an unbiased sample. Not all SNe Ia discovered during the programmes that fulfilled these criteria were observed due to scheduling constraints at VLT UT2.

The primary source of our targets was amateur searches, with discoveries taken from
the Transient Objects Confirmation Page
(TOCP\footnote{http://www.cbat.eps.harvard.edu/unconf/tocp.html}).
Additional events were discovered by the Palomar Transient Factory
\citep[PTF;][]{2009PASP..121.1334R,2009PASP..121.1395L}, La Silla
Quest Variability Survey \citep[LSQ;][]{2012Msngr.150...34B}, the
Catalina Real-Time Transient Survey
\citep[CRTS;][]{2009ApJ...696..870D}, the Lick Observatory Supernova
Search \citep[LOSS;][]{2011MNRAS.412.1419L}, and the T\'elescope \`a
Action Rapide pour les Objets Transitoires
\citep[TAROT;][]{2008PASP..120.1298K}. 

Some of the SNe were classified as part of the ESO Large Programme, Public ESO Spectroscopic Survey of Transient Objects\footnote{http://www.pessto.org/pessto/index.py} (PESSTO), which is currently operating at the New Technology Telescope, La Silla, Chile. Additional observations of SN\,2012cg are detailed in
\citet{2012ApJ...756L...7S} and \citet{2013NewA...20...30M}. Nearly all of the SNe Ia in the sample are spectroscopically similar to `normal' SNe Ia, with the exception of SN\,2013ao \cite[spectroscopically similar to a `super-Chandrasekhar' SN Ia;][]{2006Natur.443..308H}, SN\,2013U (classified as a 1991T-like object) and LSQ12gdj, a second `super-Chandrasekhar'
SN Ia (Scalzo et al. in preparation). SN\,2013ao is excluded from further discussion because a spectroscopic redshift for its host galaxy could not be determined, and hence the velocity measurements are less reliable.

\begin{table*}
  \caption{Intermediate resolution XShooter spectral information and derived light curve properties. The sample is split based on the presence of `blueshifted', `no blueshifted' or no narrow absorption features of \Nai\ D in their spectra, as discussed in Section \protect  \ref{sec:spectr-meas}. }
 \label{tab:spec}
\begin{tabular}{@{}lccccccccccccccccccccccccccccc}
  \hline
  \hline
 SN name&Date&MJD$^1$ & Phase$^2$ &MJD of $B$& Stretch&\textit{B-V}&LC source$^3$& \Nai\ D$_2$& `Blueshifted'\\
&of spec.&of spec.&(d)&-band max.&&at max.& & pEW (\AA)& \Nai\ D$_2$ pEW (\AA)$^4$ \\
\hline
\hline
&&&&&\textbf{Blueshifted \Nai\ D}\\
\hline
LSQ12fxd&20121113&56244.2&-1.9&56246.1$\pm$0.1&1.121$\pm$0.016&--&LSQ&0.50$\pm$0.01&0.50$\pm$0.01\\
LSQ12gdj	&20121118&56249.0&-4.4&56253.4$\pm$0.1&1.130$\pm$0.008&0.00$\pm$0.01&Scalzo et al. &0.05$\pm$0.03&0.05$\pm$0.03\\
SN\,2012cg&20120603&56081.0&-0.8&56081.8$\pm$0.4&1.098$\pm$0.022&0.14$\pm$0.04  &LT+RATCam$^5$&0.96$\pm$0.02&0.62$\pm$0.05\\
&20120630&56109.0&+27.3&56081.8$\pm$0.4&1.098$\pm$0.022&0.14$\pm$0.04    &LT+RATCam$^5$&0.96$\pm$0.02&0.62$\pm$0.05\\
SN\,2012et& 20120930&56201.1 &+11.1&56190.0$\pm$1.1&1.194$\pm$0.098&0.16$\pm$0.03&LT+IO:O, P48&0.65$\pm$0.04&0.65$\pm$0.04\\
SN\,2012hd$^*$	&20121127&56258.1&-6.7&56264.8$\pm$0.8&0.957$\pm$0.095&-- &FTS, LSQ&1.02$\pm$0.03&0.54$\pm$0.03\\
SN\,2013U & 20130212& 56336.2&-1.6&56337.8$\pm$0.3&1.091$\pm$0.070&--&LT+IO&0.88$\pm$0.02&0.68$\pm$0.03\\
SN\,2013aj&20130310&56360.3&+0.8&56359.5$\pm$0.6&0.911$\pm$0.054&0.02$\pm$0.02&LT+IO:O, SMARTS&0.21$\pm$0.01&0.21$\pm$0.01\\
\hline
 &&&&&\textbf{No blueshifted \Nai\ D}\\
\hline
LSQ12fuk$^*$ &20121106&56237.2&+4.6&56232.6$\pm$0.2&0.999$\pm$0.051&--&P48&0.24$\pm$0.02&--\\
SN\,2012hr&20121222&56283.1&-6.1&56289.2$\pm$0.1& 1.018$\pm$0.021&0.03$\pm$0.01&LCOGT 1m&0.12$\pm$0.01&--\\
SN\,2012fw&20120904&56174.1&+7.2&56166.9$\pm$0.7&1.135$\pm$0.073&0.11$\pm$0.02&FTS&0.24$\pm$0.01&--\\
\hline
&&&&&\textbf{No \Nai\ D}\\
\hline
LSQ12dbr	&20120702&56110.4&-0.3&56110.7$\pm$0.2&1.085$\pm$0.014&-0.13$\pm$0.02&LSQ, LT+RATCam&--&--\\
LSQ12hzj	&20130109&56301.3&+0.7&56300.6$\pm$0.2&0.931$\pm$0.035&--&LSQ&--&--\\
PTF12iiq 		&20120908&56178.3&-3.8&56182.1$\pm$0.3&0.925$\pm$0.028 &0.10$\pm$0.02&LT+RATCam, P48&--&--\\
PTF12jgb 	&20121009&56209.3&+5.4&56203.9$\pm$0.6&1.230$\pm$0.073&--&LT+IO:O, P48&--&--\\
SN\,2012ht	&20121231&56293.3 &-1.6&56294.9$\pm$0.2&0.877$\pm$0.022&-0.04$\pm$0.02&LT+IO:O&--&--\\
SN\,2013aa&20130223&56347.3&+3.3&56344.0$\pm$0.1&1.146$\pm$0.019&-0.05$\pm$0.01&LCOGT 1m&--&--\\
\hline
\end{tabular}
 \begin{flushleft}
 $^1$Modified Julian date.\\
 $^2$Phase with respect to \textit{B}-band maximum.\\
 $^3$Further information on the telescopes and instruments used can be found in Section \ref{sec:opt_phot}.\\
 $^4$`Blueshifted' \Nai\ D pEW refers to the integrated pseudo-equivalent-width of any \Nai\ D$_2$ absorption features that are blueshifted with respect to the defined zero velocity position.\\
$^5$LT light curve data were supplemented using data from \protect \cite{2013NewA...20...30M}.\\
$^*$SN 2012hd and LSQ12fuk are removed from our calculation of the ratio of `blueshifted' to `redshifted' absorption features in Section \ref{sec:excess} for the following reasons: SN 2012hd because it displays both `blueshifted' and `non-blueshifted' \Nai\ D absorption components and LSQ12fuk because its absorption is a single component at zero velocity.
\end{flushleft}
\end{table*}

\begin{table*}
  \caption{Low-resolution spectral information for follow-up spectra of our XShooter sample used for measuring the \SiII\ 6355 \AA\ velocities.The order of the SNe is following that of Tab.~\ref{tab:spec}.}
 \label{tab:lowres-spec}
\begin{tabular}{@{}lccccccccccccccccccccccccccccc}
  \hline
  \hline
    SN name&Date&MJD & Phase &Telescope+&Wavelength&\SiII\ 6355 \AA \\
    && &(d)&instrument$^1$&range (\AA)& vel. 10$^3$ \kms  \\
\hline
\hline
    LSQ12fxd&20121112&56244.2&-1.9&VLT+XSH&3100--24790& 10.97$\pm$0.10 \\
    LSQ12fxd&20121112&56244.3&-1.8&NTT+EFOSC2&3368--10300& 10.94$\pm$0.14 \\
    LSQ12gdj&20121122&56254.2&+0.8&NTT+EFOSC2&3368--10300& 10.83$\pm$0.10 \\
    SN\,2012cg&20120603&56081.0&-0.8&VLT+XSH&3100--24790& 10.42$\pm$0.10 \\
    SN\,2012hd	&20121206&56268.2&+3.4&NTT+EFOSC2&3368--10300& 9.73$\pm$0.10 \\
    SN\,2013U & 20130212& 56336.2&-1.6&VLT+XSH&3100--24790& 9.94$\pm$0.13 \\
    SN\,2013aj&20130310&56360.3&+0.8&VLT+XSH&3100--24790& 10.98$\pm$0.10 \\
    LSQ12fuk&20121106&56237.2&+4.6&VLT+XSH&3100--24790& 9.89$\pm$0.18 \\
    SN\,2012hr&20121223&56284.6&-4.6&FTS+FLOYDS&3150--10950&12.61$\pm$0.13\\
    SN\,2012hr&20130101& 56294.3&+5.3&NTT+EFOSC2&3368--10300& 11.05$\pm$0.12 \\
    SN\,2012fw&20120826&56166.1&-0.8&NTT+EFOSC2&3985--9315& 9.76$\pm$0.11 \\
    LSQ12dbr&20120702&56110.4&-0.3&VLT+XSH&3100--24790& 11.41$\pm$0.11 \\
    LSQ12hzj&20130109&56301.3&+0.7&VLT+XSH&3100--24790& 9.02$\pm$0.31 \\
    PTF12jgb&20121004&56205&0&Gemini-N+GMOS&3400--9450& 9.99$\pm$0.10 \\
    SN\,2012ht&20120231&56293.3&-1.6&VLT+XSH&3100--24790& 10.99$\pm$0.11 \\
    SN\,2012ht&20130101&56294.3 &-0.6&NTT+EFOSC2&3368--10300& 11.02$\pm$0.10 \\
    SN\,2013aa&20130219& 56342.7&-1.3&FTS+FLOYDS&3150--10950& 10.71$\pm$0.16 \\
    SN\,2013aa&20130223&56347.3&+3.3&VLT+XSH&3100--24790& 10.21$\pm$0.10 \\
    SN\,2013aa& 20130225&56348.4&+4.4&Gemini-S+GMOS&4200--8400& 10.19$\pm$0.14 \\
      \hline
\end{tabular}
 \begin{flushleft}
    $^1$Information on the telescopes and instruments used:\\
    VLT+XSH = VLT at Paranal, Chile with the XShooter spectrograph.\\
    NTT+EFOSC2 = New Technology Telescope, La Silla, Chile with the ESO Faint Object Spectrograph and Camera 2.  \\
    FTS+FLOYDS = Faulkes Telescope South, Siding Spring, Australia with the FLOYDS spectrograph.\\
    Gemini-N+GMOS = Gemini Telescope North, Mauna Kea, Hawaii, US, with the Gemini Multi-Object Spectrograph.\\
    Gemini-S+GMOS = Gemini Telescope South, Cerro Pachon, Chile, with the Gemini Multi-Object Spectrograph.\\
\end{flushleft}
\end{table*}

\subsection{Spectroscopy}
\label{highres_spec}

\begin{figure*}
\includegraphics[width=14.5cm]{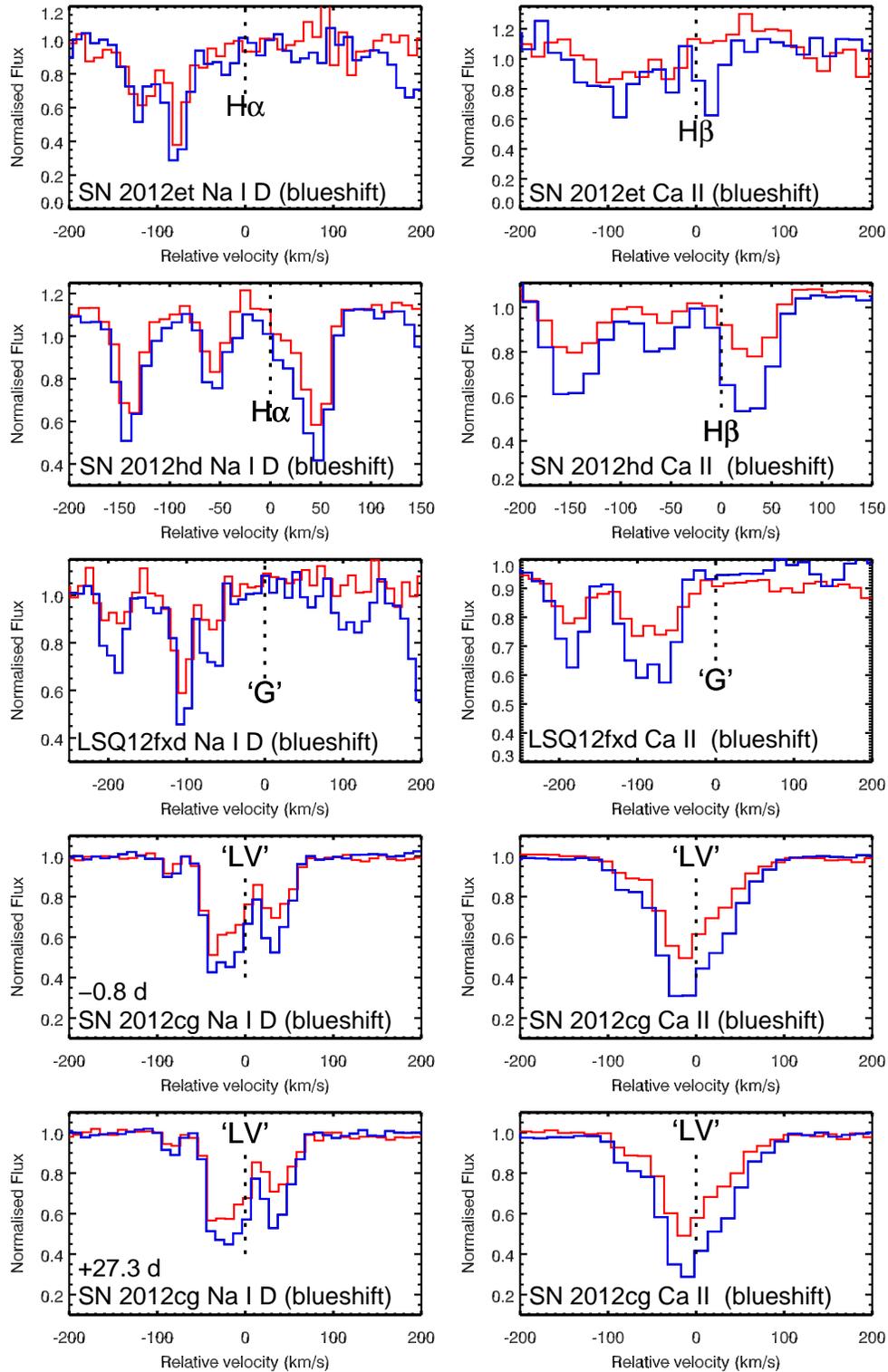}
\caption{Intermediate-resolution VLT+XShooter spectra of the \Nai\ D
  absorption lines in the left panels (\Nai\ D$_1$ is in red, \Nai\ D$_2$ is in blue) and the \Caii\ H\&K lines on the
  right (\Caii\ H is red, \Caii\ K is in blue). The velocity scale is plotted with a zero velocity defined by
  the position of the host galaxy features, marked with  `H$\alpha$'  and `H$\beta$'.  If galaxy lines are not visible, the rest wavelength is  defined by the recessional velocity of the host galaxy obtained from
  NED, and is marked with a `G' on the spectra. For SN\,2012cg, the local velocity (`LV') at the SN position was measured from the stellar velocity maps of \citet{2006AJ....131..747C}. Two spectra of SN\,2012cg were obtained at -0.8 d and +27.3 d with respect to \textit{B}-band maximum. All SNe Ia in
  this figure are labelled as having absorption features with a `blueshift'. SN\,2012hd shows both `blueshifted' and `redshifted' \Nai\ D features, which will be discussed in Section \ref{sec:spectr-meas}.}
\label{csm1}
\end{figure*}

\begin{figure*}
\includegraphics[width=14.5cm]{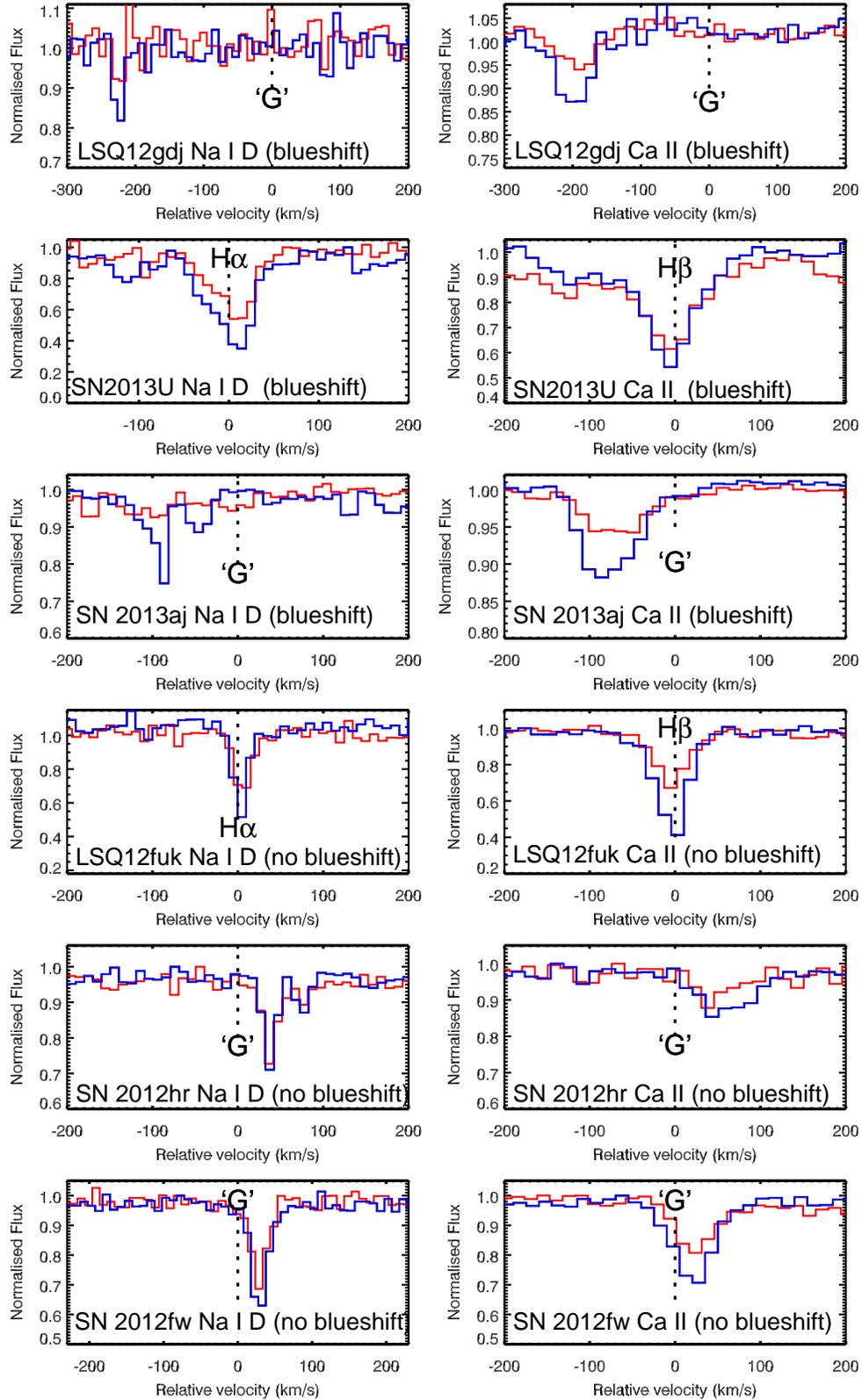}
\caption{As Fig.~\ref{csm1}. SNe Ia displaying `blueshifted' absorption features are marked as `blueshift', while those not displaying any blueshifted features are labelled as 'no blueshift'.}
\label{csm2}
\end{figure*}

\begin{figure*}
\includegraphics[width=14.5cm]{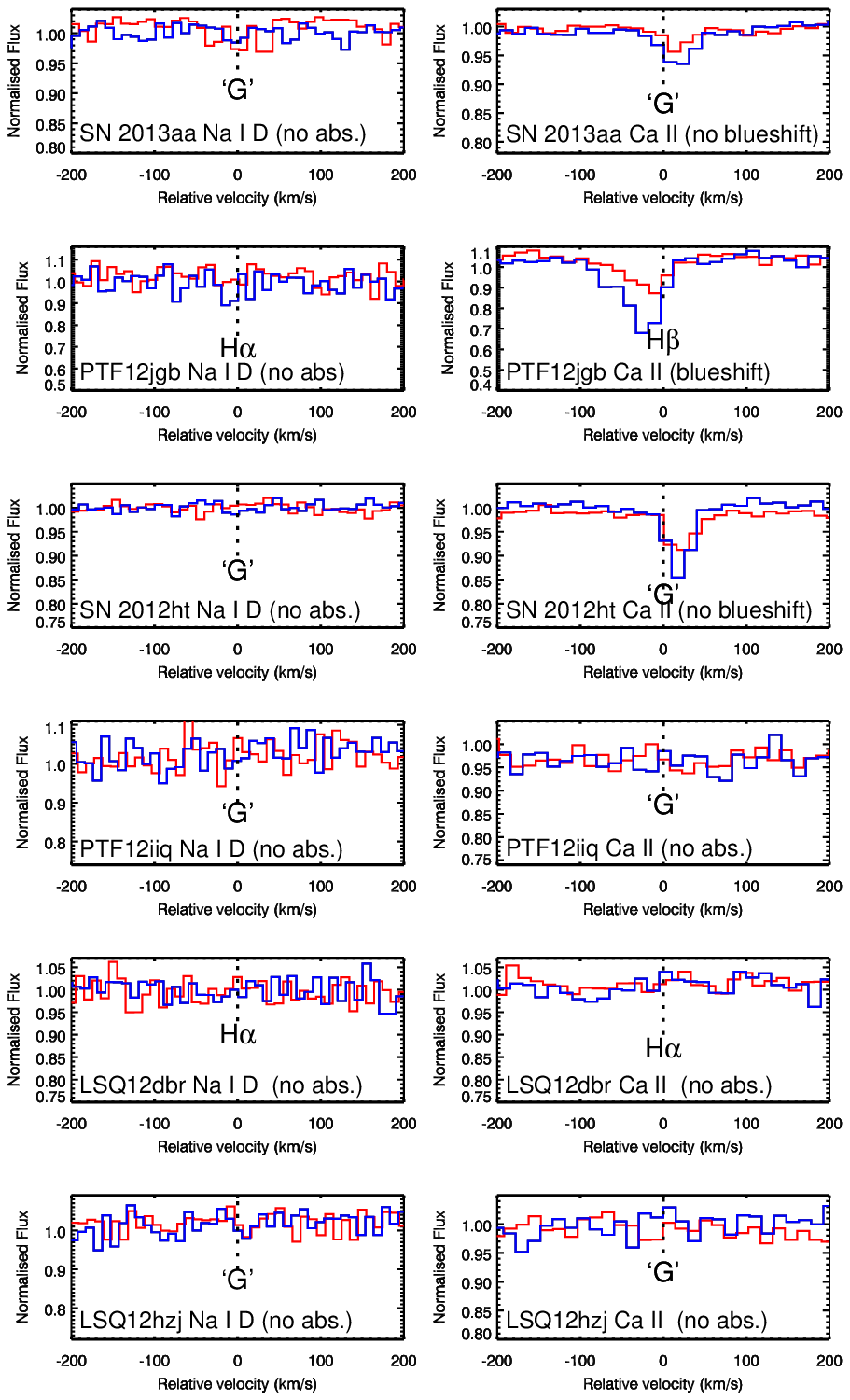}
\caption{As Fig.~\ref{csm1}. None of the SNe Ia in this figure display
 absorption features in the \Nai\ D wavelength (`no abs.'), while two show `non-blueshifted', one shows `blueshifted' and three show no \Caii\ H\&K absorption. }
\label{csm3}
\end{figure*}

Intermediate-resolution spectra were obtained for all SNe using
XShooter under target-of-opportunity (ToO) programmes, IDs
089.D-0647(A) and 090.D-0828(A). XShooter is an echelle spectrograph
with three arms (UV, visible and near-infrared) covering the
wavelength range of 3000--25000 \AA. The instrumental resolution of
the spectrograph is fixed, and we used the narrowest available slit
widths of 0.5\arcsec\ (UV arm), 0.4\arcsec\ (visible arm) and
0.4\arcsec\ (near-infrared arm) to achieve resolutions of $R\sim9900$,
$R\sim18200$ and $R\sim10500$, respectively. These resolutions are necessary to resolve the narrow \Nai\ D (5890, 5896 \AA),  \Caii\ H\&K (3934, 3969 \AA)  and \Ki\ (7665,7699 \AA) features of interest for this study. Due to the narrow slits employed and lack of atmospheric dispersion correctors for XShooter at the time of observations, the absolute flux calibration is uncertain. However, we are interested in the shifts in wavelength position and the relative strengths of features so this does not affect our results. Details of the XShooter spectral observations are listed in Table~\ref{tab:spec}. The spectra were
reduced using the public XShooter pipeline, which performs a full
reduction of the spectral orders in each of the three arms to obtain a
contiguous one-dimensional merged spectrum \citep{2010SPIE.7737E..56M}.
Spectra showing the \Nai\ D and \Caii\ H\&K narrow absorption features are displayed in Figs.~\ref{csm1} to \ref{csm3} for the whole sample. XShooter spectra of SN\,2012cg were obtained on two occasions, with both shown in Fig.~\ref{csm1}. The selection of velocity zero points for the sample is discussed in Section \ref{sec:analysis}.

Follow-up low-resolution optical spectra were obtained for our
SN sample and details are given in Table~\ref{tab:lowres-spec}. These low-resolution
spectra were reduced using custom pipelines for each of the telescopes
based on standard spectral reduction procedures in \textsc{iraf} and
\textsc{idl}. The two-dimensional spectra were bias and flat-field corrected before
extraction. The extracted spectra were calibrated in wavelength using
arc-lamp exposures and instrumental response functions were obtained
from observations of spectrophotometric standards to perform the flux
calibration. 

\subsection{Optical photometry}
\label{sec:opt_phot}

The optical photometry of the SNe Ia comes from six facilities: i)
the PTF search telescope, the Palomar 48-in (P48), ii)  the LSQ search telescope, the 40$''$ ESO Schmidt Telescope, iii) the robotic 2-m
Liverpool Telescope \citep[LT;][]{2004SPIE.5489..679S}, iv) the Faulkes
Telescope South (FTS), v) Las Cumbres Observatory Global Telescope
(LCOGT) 1-m telescope array in Chile, part of the LCOGT network \citep{2013arXiv1305.2437B}, and vi) the SMARTS 1.3m telescope at Cerro Tololo Inter-American Observatory (CTIO), Chile. \textit{g'} and \textit{R}-band data were taken with the P48, and reduced using the Infrared
Processing and Analysis Center
(IPAC)\footnote{http://www.ipac.caltech.edu/} pipeline (Laher et
al.~in preparation) and photometrically calibrated \citep{2012PASP..124..854O}.
 The LSQ telescope observes in a wide `\textit{g+r'} filter, while the LT data were obtained using both the RATCam
and IO:O optical imagers in \textit{gri} filters, similar to those used in
the Sloan Digital Sky Survey \citep[SDSS;][]{2000AJ....120.1579Y}. Data from FTS, a clone of the LT, were obtained with the SPECTRAL imager in $gri$ filters, while data from the LCOGT 1-m array were obtained in Johnson-Cousins \textit{UBVRI} and SDSS-like \textit{gri} filters. SMARTS data were obtained using the optical-infrared imager, ANDICAM in KPNO $BVRI$ filters.  Photometric data of superluminous SN Ia, LSQ12gdj will be presented in Scalzo et
al.~in preparation and is used here to calculate a light curve width and colour.

The SN magnitude was measured on each epoch using PSF photometry and calibrated using tertiary standard stars in the field of the SN. Where available, the zero points of the images were calculated using aperture photometry by comparing tertiary stars in the field directly with their SDSS magnitudes. If this was not possible, stars in the SN fields were calibrated using standard star observations on photometric nights and used to estimate a nightly zero point. Uncertainties on the SN flux measurements are a combination of the statistical and calibration uncertainties and are inputted to the light curve fitting routine, detailed in Section \ref{sec:light-curve-fitting}.

\subsection{Light curve fitting}
\label{sec:light-curve-fitting}
The optical light curves were analysed using the SiFTO
light curve fitting code \citep{2008ApJ...681..482C}, which produces values for the
stretch ($s$; light curve width), maximum $B$-band magnitude, $B-V$ at maximum (where multiple bands are available), and time of
maximum light for each SN.  These values for our sample are given in Tab.~\ref{tab:spec}. SiFTO uses a time-series of spectral
templates that are adjusted to recreate the observed colours of the SN
photometry at each epoch, while also adjusting for Galactic extinction
and redshift (i.e., the $k$-correction). The unknown contribution of the host galaxy to extinction has not been corrected for. All of the SNe Ia in our sample fall within the light curve width range used in cosmological studies ($0.7<s<1.3$).

\subsection{SN photospheric \SiII\ velocity measurements}
\label{sec:velo}

SNe Ia with `blueshifted' \Nai\ D absorption features (with respect to the strongest \Nai\ D component) have been suggested to have higher \SiII\ 6355 \AA\ velocities near maximum light relative to the SN Ia population as a whole \citep{2012ApJ...752..101F}.  Therefore, we measure the velocities of \SiII\ 6355 \AA\ feature in near-maximum light spectra for our sample.

The  \SiII\ 6355 \AA\  velocities were measured using a Gaussian fit to the
feature, broadly following the method described in
\cite{2012MNRAS.426.2359M}. Briefly, the pseudo-continuum is defined using a region on each side of the feature and then the observed spectrum is divided by this continuum. The position of the minimum of the feature is measured by fitting
a Gaussian to the feature using the \textsc{mpfit}
\citep{2009ASPC..411..251M} procedure in \textsc{idl}. Both the value
of the pseudo-continuum and the wavelength range used in the fit are
varied to obtain the mean velocity of the feature along with the
uncertainties in these measurements. Redshift uncertainties are also included in quadrature. 

\citet{2011ApJ...742...89F} and \citet{2012MNRAS.425.1819S} make independent calculations to correct \SiII\ 6355 \AA\ velocities to maximum light but find inconsistent relationships between velocity and phase. The relationships are also only accurate in a narrow light curve width range, 1$<\Delta m _{15}(B)$\footnote{The decay of the \textit{B} band magnitude between maximum and 15 d post-maximum.}$<$1.5 mag. Since some of our SNe Ia are not in this light curve width range, we chose not to correct to maximum light and instead limit our analysis to a small phase range of -2 to +5 d with respect to maximum to minimise time-dependent variations. Not correcting the spectra to maximum light using the calculations of  \citet{2012MNRAS.425.1819S}, results in additional uncertainties of $<$200 \kms\ for our sample. For SNe Ia with spectra before and after maximum light (but limited to the range -5 to +5 d), we use a linear fit to estimate the value at maximum light. The \SiII\ 6355 \AA\ velocities are given in Tab.~\ref{tab:lowres-spec}.

\subsection{\Nai\ D pseudo equivalent-width measurements}
\label{sec:pseudo-equiv-width}

The pseudo-equivalent width (pEW)\footnote{These are pEW and not true pEW since the continuum of a SN Ia spectrum is not a `true' continuum but made up of many blended absorption lines} of the narrow \Nai\ D$_2$ absorption feature are also measured to investigate the potential contribution of CSM to the strength of the absorption. We use the \Nai\ D$_2$ feature since it is the stronger of the two \Nai\ D lines.  We calculate the pEW of the \Nai\ D$_2$ components by firstly selecting the continuum on either side of the features of interest and then computing the area below the continuum, similar to the method of \cite{2012ApJ...754L..21F}. The pEW for the individual components are then totalled to provide an estimate of the pEW of the \Nai\ D$_2$ line. This pEW measurement will contain contributions from host galaxy interstellar absorption, as well as potentially CSM from the SN progenitor system if present. 

\subsection{Literature data}
\label{sec:literature-data}

We supplement our SN Ia sample with events from the sample of \citet{2011Sci...333..856S}. They chose their high-resolution spectral sample to only include SNe Ia that were not specifically targeted due to either strong \Nai\ D lines in classification spectra or showing red colours, which could be indicative of a CSM contribution, and which would have resulted in a biased sample. Since our data were similarly selected, we exclude other samples of high-resolution spectra if their selection criteria are not clearly outlined. This removes any potential bias from SNe Ia displaying positive detections of narrow absorption features being preferentially published.

We have collated photometric data from the literature for 16 of the 35 SNe Ia of \citet{2011Sci...333..856S}: SN\,2006X, SN\,2006cm,  SN\,2007af, SN\,2007kk, SN\,2009ds, SN\,2009ig, SN\,2010A from \citet{2009ApJ...700..331H,2012ApJS..200...12H}, SN\,2007le, SN\,2007on,  SN\,2008C, SN\,2008fp,  SN\,2008hv, SN\,2008ia from \citet{2011AJ....142..156S}, SN\,2008ec, SNF20080514-002 from \citet{2010ApJS..190..418G} and SN\,2009le from
\citet{2012MNRAS.426.2359M}. The light curve width and colour of these SNe were fit using SiFTO. The rest of the SNe Ia in the sample did not have well constrained light curve fits, due to sparse data coverage or poor data quality, or had stretch values outside the range used in cosmological studies ($0.7<s<1.3$).  

We have gathered near-maximum light spectra of the following SNe Ia from the
literature: SN\,2006X \citep{2008ApJ...675..626W}, SN\,2006cm, SN\,2007le, SN\,2007kk, SN\,2008C, SNF20080514-002
\citep{2012AJ....143..126B}\footnote{Spectra obtained from the Weizmann
  Interactive Supernova data Repository (WISeREP) --
  www.weizmann.ac.il/astrophysics/wiserep;
  \citep{2012PASP..124..668Y}}, SN\,2007af \citep{2012AJ....143..126B,2007ApJ...671L..25S} SN\,2009ig \citep{2012ApJ...744...38F}
and SN\,2009le \citep{2012MNRAS.426.2359M}. Maximum light spectra
of SN\,2007on, SN\,2008fp, SN\,2008hv and SN\,2008ia were obtained by
the Carnegie Supernova Project (Mark Phillips, priv. communication).  This sample of 16 SNe Ia from \citet{2011Sci...333..856S} with calculated light curve parameters will be referred to as the S11 sample hereafter. The light curve parameters and \SiII\ 6355 \AA\ velocities are given in Table \ref{tab:appda}.

\section{Analysis}
\label{sec:analysis}

We now turn to the analysis of our sample. We
discuss first the definition of the zero velocity of the narrow absorption features, critical for defining relative velocity offsets. We then detail how the properties of the narrow absorption feature sample is dependent on host galaxy, light curve and spectral properties. 

\subsection{Spectral measurements of narrow absorption features}
\label{sec:spectr-meas}

For our analysis of the XShooter spectra (Figs.~\ref{csm1} to
\ref{csm3}), we are interested in the position of the very narrow absorption features of \Caii\ H\&K and
\Nai\ D relative to a defined zero velocity
(rest-frame). These features do not originate from material in the SN ejecta, but rather from absorbing material along the line-of-sight, which can be caused by ISM in the host galaxy and/or CSM around the SN progenitor. 

The frequency of occurrence of each type of profile with respect to a defined zero velocity
(i.e., blueshifted, non-blueshifted or no absorption) can be measured,
and the relative rates and possible correlations with other
observables can then be determined. Hence the definition of `zero
velocity' is critical, and can be set in a number of ways.

Previous studies, such as \citet{2011Sci...333..856S} and
\citet{2012ApJ...752..101F}, chose the strongest narrow \Nai\ D absorption
component as a proxy for the galaxy component at the SN position.
`Blueshifted', `redshifted', `single/symmetric' or `no absorption' features
are then determined relative to this position \citep[see][for further information]{2011Sci...333..856S}. One disadvantage of
this method is in cases where no host galaxy ISM absorption exists
(perhaps more likely in early-type galaxies), SN Ia CSM features
will then be misidentified as `single/symmetric' even though they may
originate from the SN environment itself.

\begin{table}
 \caption{Classification of \Nai\ D features for XShooter sample and combined sample with S11.}
 \label{tab:class}
 \begin{tabular}{@{}lccccccccccccccccccccccccccccccc}
  \hline
  \hline
\Nai\ D classification&XShooter sample&Combined sample\\
\hline
Blueshifted only&6& 11\\
Redshifted only&2&4\\
Blue and Redshifted&1&6\\
Symmetric&1&1\\
None&6&10\\
Excess of blueshifted$^1$&25\%&22\%\\
\hline
Total& 16&32\\
\hline
 \end{tabular}
\begin{flushleft}
 $^1$Excess of SNe Ia with only blueshifted \Nai\ D absorption features compared to only redshifted \Nai\ D absorption as a percentage of the total sample.\\
\end{flushleft}
\end{table}

Our method is to set the zero velocity relative to the
positions of galaxy lines (e.g., narrow nebular emission lines of H$\alpha$ and H$\beta$) in the SN Ia spectrum. This probes the rest-frame velocity along the
line-of-sight to the SN position. If these lines are not visible, the
rest wavelength is instead  set using the recessional velocity of
the host galaxy, taken from the NASA/IPAC Extragalactic Database
(NED) or SDSS. The advantage of this method is that
`single/symmetric' profiles can now be assigned a blueshifted or
non-blueshifted position relative to the other host galaxy lines. The
disadvantage is that using the recessional velocity does not account
for any internal motion or rotation in the host galaxy. To estimate this potential offset for our SN Ia sample, we compare the SNe Ia that have a velocity measured from NED/SDSS and host galaxy lines in the spectra. The differences range from 20 \kms\ for SN\,2013U to 150 \kms\ for SN\,2012hd. When using the recessional velocities instead of host galaxy lines, one SN moves from being classified as `blueshifted' to `non-blueshifted' (SN\,2012hd\footnote{SN\,2012hd has multiple \Nai\ D and \Caii\ H\&K absorption features that span a wide range of velocities. If the velocity of the SN progenitor system is assumed to be equal to that of the host galaxy emission lines then these features appear at both blueshifted and redshifted velocities, while if the recessional velocity of the host galaxy is used instead, all of the absorption features are redshifted.}).  If the observed \Nai\ D and \Caii\ H\&K absorption features are all due to the host galaxy, one could expect an equal number of SNe Ia showing blueshifted and non-blueshifted absorption features. However, if there is an additional contribution to the blueshifted sample from the CSM, an excess of blueshifted features is expected. As this is a statistical argument, we do not discuss the CSM properties of individual SNe Ia in our sample. The zero velocity positions of the S11 sample were also reanalysed for this study. The wavelength regions around their \Nai\ D features were downloaded from WISeREP.

Previous studies have only identified time-varying blueshifted \Nai\ D absorption features, while the \Caii\ H\&K features have been found not to vary due to its higher ionisation potential \citep{2007Sci...317..924P,2009ApJ...702.1157S}. Indeed the non-detection of varying \Caii\ H\&K is central to the argument for a CSM origin to the \Nai\ D variations \citep[see][for detailed discussion]{2007Sci...317..924P}. Therefore, we perform our analysis of velocity shifts of narrow absorption features using the \Nai\ D feature only. However, we show the \Caii\ H\&K region in Figs.~\ref{csm1} to
\ref{csm3}, where it is seen that classification of the  \Nai\ D and \Caii\ H\&K features (and relative velocities) agree for our XShooter sample for all but three SNe. The \Caii\ H\&K region is also useful for confirming weak line detections at \Nai\ D wavelengths. 

We also searched for narrow \Ki\ absorption features. These features are expected
to be much weaker than those of \Caii\ H\&K and \Nai\ D. SN\,2012fw was
the only SN in the sample to display \Ki\ absorption and showed
redshifted features, as did the \Caii\ H\&K and \Nai\ D absorption features for this
event.

\subsection{Defining the relative velocities}
Having defined a zero velocity position for each SN, we then search for absorption features near the zero velocities of the \Nai\ D and \Caii\ H\&K lines. We wish to extend the statistical analysis of \citet{2011Sci...333..856S}, and estimate the ratio of SNe Ia with blueshifted to redshifted \Nai\ D absorption profiles in our sample. To do this, we classify our SN sample into five separate \Nai\ D absorption categories: i) SNe Ia with just blueshifted absorption, ii) SNe Ia with just redshifted absorption, iii) SNe Ia with blueshifted and redshifted features, iv) SNe Ia with single absorption features at zero velocity, and v) SNe Ia with no \Nai\ D absorption features. We split the sample in this way so that SNe Ia with only blueshifted absorption \Nai\ D absorption features can be directly compared to those with only redshifted \Nai\ D absorption features, as was done in \citet{2011Sci...333..856S}.

The classifications of the \Nai\ D features for both the XShooter sample and the combined sample with S11 are given in Table \ref{tab:class}. Six SNe show both blueshifted and redshifted features: SN\,2006cm, SN\,2007af, SN\,2009ds, SN\,2009le, SN\,2010A, SN\,2012hd, while LSQ12fuk is classified as symmetric since it shows only a single component at zero velocity with respect to H$\alpha$. The analysis of the ratio of SNe Ia in our sample with blueshifted to redshifted \Nai\ D absorption features is detailed in Section \ref{sec:excess}.

For our analysis of the light curve width and host galaxy properties as a function of \Nai\ D absorption profile properties, we define our \Nai\ D absorption categories by splitting the sample into three: i) SNe Ia with any blueshifted absorption profiles (`blueshifted'), ii) SNe Ia with only non-blueshifted absorption profiles (`non-blueshifted') and iii) SNe Ia with no \Nai\ D absorption features (`no absorption'). This different classification is made since SNe Ia with `non-blueshifted' \Nai\ D absorption features will always be associated with the host galaxy as they are not related to outflowing material and can not be associated with the progenitor system. Conversely, SNe Ia displaying both blueshifted and redshifted \Nai\ D absorption features are now included in the `blueshifted' category since any blueshifted \Nai\ D absorption profiles seen \textit{may} be associated with progenitor outflow -- even though we can not determine if this is the case for any individual SN. The breakdown of individual SNe Ia from our XShooter sample into these \Nai\ D absorption categories is given in Table \ref{sec:spectr-meas}, while information on the SNe Ia from the S11 sample is given in Table \ref{tab:appda}.

In Section \ref{sec:colvel}, we compare our colour and spectra analysis to those of \cite{2012ApJ...752..101F}, who split their sample into two categories, those that have blueshifted \Nai\ D absorption features and those that do not (`everything else'). We also use these categories for investigating the pEW of \Nai\ D features in \ref{sec:relstrength}, since we wish to determine if SNe Ia with blueshifted \Nai\ D absorption features have stronger pEW than the rest of the sample. 

\subsection{Statistics of \Nai\ D absorption profiles}
\label{sec:excess}

 \begin{figure}
\includegraphics[width=8.5cm]{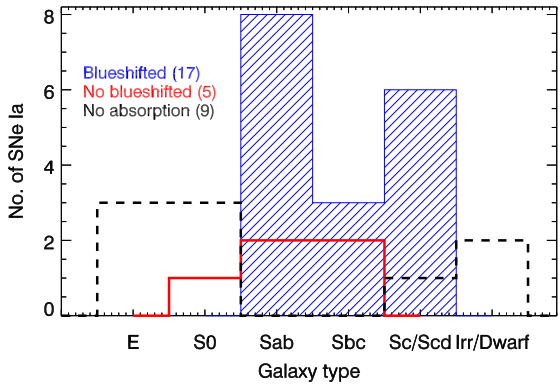}
\caption{Histogram of the host galaxy types of 31 SNe Ia  from the
  combined S11 sample and our XShooter sample, colour-coded based on the presence of
  `blueshifted' (blue hashed region; any blueshifted \Nai\ D features including those showing blueshifted and non-blueshifted features), `no blueshifted' (red solid line; any non-blueshifted absorption features)  or `no absorption' (black dashed line; no \Nai\ D absorption features).  }
\label{csm_gal1}
\end{figure}

\citet{2011Sci...333..856S} identified an excess of blueshifted over redshifted \Nai\ D absorption profiles in their SN Ia sample (12 SNe with `blueshifted' features, 5 with `redshifted' features, 5 with `symmetric' profiles and 13 showing no narrow \Nai\ D absorption features). 

For the Xshooter sample alone, we find more SNe Ia showing blueshifted (6) \Nai\ D absorption features compared to redshifted  (2) \Nai\ D absorption features (see Table \ref{tab:spec} for the information on the individual SNe Ia and Table \ref{tab:class} for a summary of the total numbers in each grouping for this comparison.) Assuming an equal probability of obtaining a \Nai\ D absorption profile that is `blueshifted' or `redshifted', we find a 14 per cent chance of this result occurring at random. For the combined sample, we also find more SNe Ia with only blueshifted \Nai\ D absorption features (11) compared to only redshifted \Nai\ D absorption features (4) for the combined sample, which has a 6 per cent chance of occurring at random. If equal numbers of the SNe Ia with blueshifted and redshifted \Nai\ D absorption features are assumed to have a host galaxy ISM origin, this gives an excess of 7 SNe Ia with anomalous blueshifted \Nai\ D features. This is equivalent to 22 per cent of the total sample of 32 SNe Ia having an additional blueshifted component. 

For comparison, we also use the original method of \citet{2011Sci...333..856S} for our combined sample (where the strongest component is chosen as the velocity zero point) and find 14 SNe with blueshifted features, 4 SNe with redshifted features, 3 with symmetric features and 10 with no absorption features -- again showing an excess of blueshifted over redshifted features  ($\sim$30 per cent of the sample) with 1.5 per cent chance occurrence probability.

 \begin{figure}
\includegraphics[width=9cm]{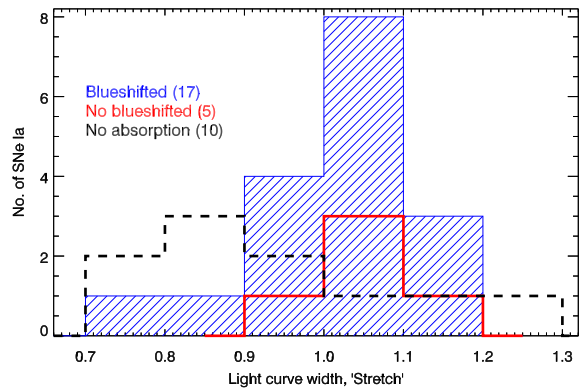}
\caption{Histogram of the light curve width parameter, `stretch', colour-coded depending on the presence of blueshifted \Nai\ D absorption features (blue hashed region), no blueshifted \Nai\ D absorption features (red solid line) or no \Nai\ D absorption (black dashed line).  }
\label{csm_stretch}
\end{figure}

\subsection{Host galaxy distribution}
\label{sec:host-galaxy-distr}

Fig.~\ref{csm_gal1} shows the morphological type distribution for the combined sample of our data and the S11 sample, split into those that show blueshifted (any SN Ia showing any blueshifted \Nai\ D features),
no blueshifted (containing `non-blueshifted' absorption features) or no \Nai\ D absorption features. PTF12jgb is excluded here because its galaxy morphology could not be determined. The galaxy categories are `E', `S0', `Sab', `Sbc', `Sc/Scd', and `Irregular/Dwarf'. The host galaxies of two SNe Ia in the sample are categorised as S0/a (SN\,2012fw, SN\,2008fp). We classify them half each in the `S0' and `Sa' bins but because they are both classified as `non-blueshifted' this does not result in non-integer values in Fig.~\ref{csm_gal1}.

We find that SNe Ia with detected narrow \Nai\ D absorption features are found more frequently in late-type galaxies, as has been found in previous studies \citep{2012ApJ...752..101F,2013ApJ...772...19F}. This is expected as star-forming galaxies typically contain denser ISM. Conversely, the SNe Ia displaying no \Nai\ D features are predominantly found in early-type host galaxies. No SN Ia with blueshifted \Nai\ D features in our sample is found in an early-type galaxy (E/S0) compared to 17 `blueshifted' SNe Ia in later-type hosts. From Section \ref{sec:excess}, we estimate that $\sim$11 of the SNe Ia displaying blueshifted \Nai\ D features are caused by CSM, with the remaining $\sim$6 due to ISM\footnote{If equal numbers of the SNe Ia with blueshifted and redshifted only \Nai\ D absorption features are assumed to have a host galaxy ISM origin, then four out of the 11 showing blueshifted only \Nai\ D features should be due to ISM, which is $\sim$36 per cent. Then 36 per cent of the 17 SNe Ia showing `blueshifted only' or `blueshifted and redshifted' features is equal to 6 SNe Ia in the combined `blueshifted' sample having features due to ISM, with the rest (11 SNe Ia) caused by CSM.}. However, we can not determine which SNe Ia fall into which category since we do not have the necessary information for any individual event. For the two SNe in elliptical host galaxies, neither show narrow \Nai\ D absorption features.

\subsection{Light curve width}

Given the excess of blueshifted SNe Ia in late-type galaxies, the lack of blueshifted SNe Ia in early-type hosts, and the well-studied connection between light curve width and host galaxy properties \citep[SNe Ia in later-type galaxies having broader light curves or higher `stretch' than those in early-type galaxies;][]{1995AJ....109....1H,1996AJ....112.2391H,1999AJ....117..707R,2000AJ....120.1479H}, we wish to investigate potential correlations between the presence of blueshifted \Nai\ D absorption profiles and SN Ia luminosity (using light curve width as a proxy). 

Fig.~\ref{csm_stretch} shows a histogram of the light curve width parameter, `stretch', for our SN sample, colour-coded  based on the presence of blueshifted \Nai\ D absorption features, `non-blueshifted' absorption features or no \Nai\ D absorption. SNe Ia displaying \Nai\ D absorption features have, on average, broader light curves than those with no \Nai\ D absorption. A Kolmogorov-Smirnov (K-S) test shows that there is a high probability (p-value=0.015) that the `blueshifted' and `no blueshifted' stretch distributions are drawn from a different parent population to the `no absorption' sample's stretch distribution.  This is not surprising since the SNe Ia with no \Nai\ D absorption are located preferentially in early-type galaxies, where SNe Ia are well known to have narrower light curves than those in later-type host galaxies. We note that any K-S test probability involving the `blueshifted' sample is actually a lower limit, since we expect that $\sim$6 SNe Ia in the `blueshifted' sample are not due to ISM (not CSM) in the host galaxy and therefore, contaminate the `blueshifted' population.

Within the subsample of SNe Ia showing non-zero \Nai\ D absorption features, we find a low probability of the `blueshifted' and `non-blueshifted' stretch distributions being drawn from different parent populations, with both samples displaying broader light curves, and occurring more frequently in late-type galaxies than those with no \Nai\ D absorption features.

 \subsection{SN colour and \SiII\ 6355 \AA\ velocity}
 \label{sec:colvel}
 
\begin{figure}
\includegraphics[width=9cm]{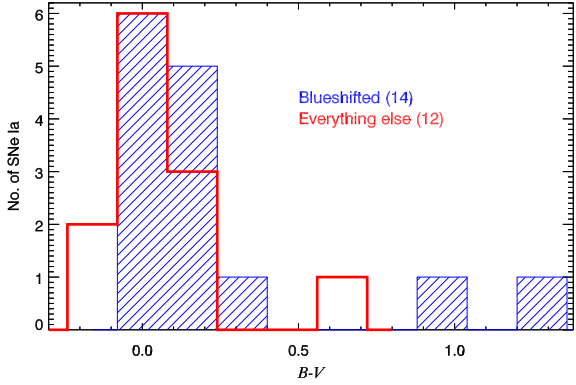}
\caption{ $B-V$ colour distribution of the SN Ia sample in all galaxy types, colour-coded based on the presence of blueshifted \Nai\ D absorption features (blue hashed) or absence (`everything else', red line). The bin size of 0.16 is chosen as twice the largest uncertainty on a colour measurement. The numbers in parentheses indicate the number of SNe Ia in each \Nai\ D group.}
\label{colour_stretch}
\end{figure}

\begin{figure}
\includegraphics[width=9cm]{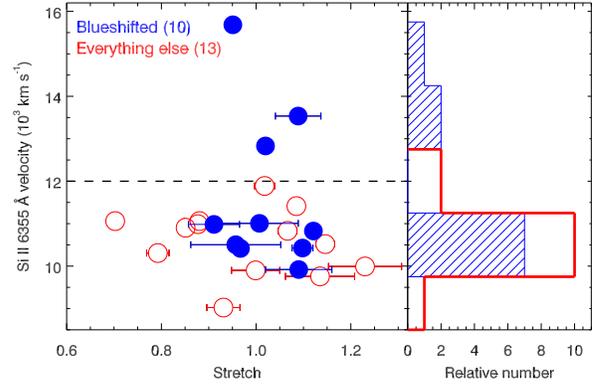}
\caption{The \SiII\ 6355 \AA\ velocity around maximum light is shown against the light curve width parameter, stretch. The SNe are colour-coded based on the presence (blue, solid circles) or absence (red, open circles) of blueshifted \Nai\ D absorption profiles. The uncertainties on the velocity measurements are smaller than the symbols. The dashed horizontal line marks the lower limit for `high-velocity' SNe Ia as defined by \protect \cite{2013arXiv1303.2601W} of 12000 \kms.}
\label{si_stretch_csm}
\end{figure}

Claims that SNe Ia displaying blueshifted \Nai\ D absorption features have redder optical colours ($B_{max}-V_{max}$ pseudo-colour\footnote{\textit{B}-band magnitude at \textit{B} maximum minus \textit{V} magnitude at \textit{V}-band maximum.}) and having increased \SiII\
6355 \AA\ line velocities than the rest of the SN Ia population have been made \citep{2012ApJ...752..101F}. Firstly, we investigate the relationship between SN colour and blueshifted \Nai\ D features. We choose to study the $B-V$ colour at the time of \textit{B}-band maximum since it is a more physical quantity than $B_{max}-V_{max}$ pseudo-colour. However, the two quantities are strongly correlated  \citep{2012AJ....143..126B} and our choice does not affect the results.

To make a suitable comparison to \cite{2012ApJ...752..101F}, we group the SNe Ia in our sample into those displaying blueshifted \Nai\ D (`blueshifted') and those that do not, including both those that show non-blueshifted \Nai\ D absorption and those with no absorption features (`everything else'). In Fig.~\ref{colour_stretch}, we show the $B-V$ colour distributions for the full SN Ia sample. A K-S test for the samples gives a p-value of 0.10 that the `blueshifted' and `everything else' samples are drawn from different parent $B-V$ colour populations, which is larger than the typically used significance level of $<$0.05. Any difference is primarily driven by a single object SN\,2006X, which had an unusually red $B-V$ colour of 1.22$\pm$0.01, making it an outlier to the SN population as a whole. 

The connection between \SiII\ 6355 \AA\ velocity and blueshifted \Nai\ D absorption profiles is also investigated. Fig.~\ref{si_stretch_csm} shows the maximum light \SiII\ 6355 \AA\ velocity as a function of stretch for both the `blueshifted' and `everything else' samples. Using a K-S test, we find a very low probability (p-value=0.7) that the `blueshifted' and `everything else' samples are drawn from different parent \SiII\ velocity distributions. We note that the three SNe Ia (SN\,2006X, SN\,2007le, SN\,2009ig) with the highest  \SiII\ 6355 \AA\ velocities do show blueshifted \Nai\ D features and fall in the `high-velocity' (HV) class of SNe Ia \citep{2009ApJ...699L.139W}, defined as having a \SiII\ 6355 \AA\ velocity of $>$12000 \kms. However, the velocity distribution of the blueshifted sample does not show a statistically significant difference from the rest of our SN Ia sample. 

\subsection{Relative strength of \Nai\ D absorption components}
\label{sec:relstrength}

The pEW of the \Nai\ D features can be used to estimate the relative amount of absorbing material along the line-of-sight towards the SNe in the sample. We do not attempt to convert these values to column densities but perform a relative comparison between SNe Ia in our sample showing blueshifted material and the `everything else' sample as defined in \ref{sec:colvel}. 

\begin{figure}
\includegraphics[width=9cm]{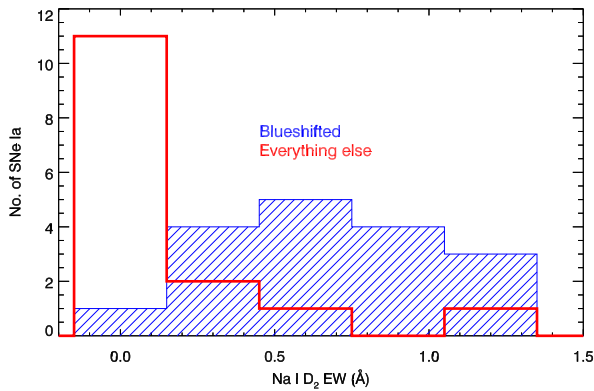}
\caption{Histogram of the pEW of the narrow \Nai\ D$_2$ features of the SN Ia sample for all SNe in the sample, colour-coded depending on the
  presence (blue hashed) or absence (red line) of blueshifted \Nai\ D absorption features.}
\label{naid_ew}
\end{figure}

Fig.~\ref{naid_ew} shows the sum of the pEW of \Nai\ D$_2$ components for each SN Ia split based on the presence or absence of blueshifted \Nai\ D absorption profiles. The \Nai\ D$_2$ pEW values are given in Table \ref{tab:spec} for the XShooter sample and in Table \ref{tab:appda} for the S11 sample. We find that SNe Ia with blueshifted \Nai\ D absorption features have higher \Nai\ D$_2$ pEW values than those without blueshifted \Nai\ D absorption features. A K-S test gives a very high probability (p-value=0.0004) of the `blueshifted' and `everything else' \Nai\ D$_2$ pEW distributions being drawn from different parent populations. 

These higher \Nai\ D$_2$ pEW in the `blueshifted' sample may potentially explained by an additional contribution from CSM to their strength. To investigate the origin of this additional contribution further, we measure the integrated pEW of only the  `blueshifted'  \Nai\ D$_2$ components (excluding any redshifted components) in the `blueshifted' SN Ia sample. The `blueshifted' \Nai\ D$_2$ pEW is then defined as the pEW of any features (fractional or full) that are blueshifted with respect to our defined zero velocity. 

\begin{figure*}
\includegraphics[width=18cm]{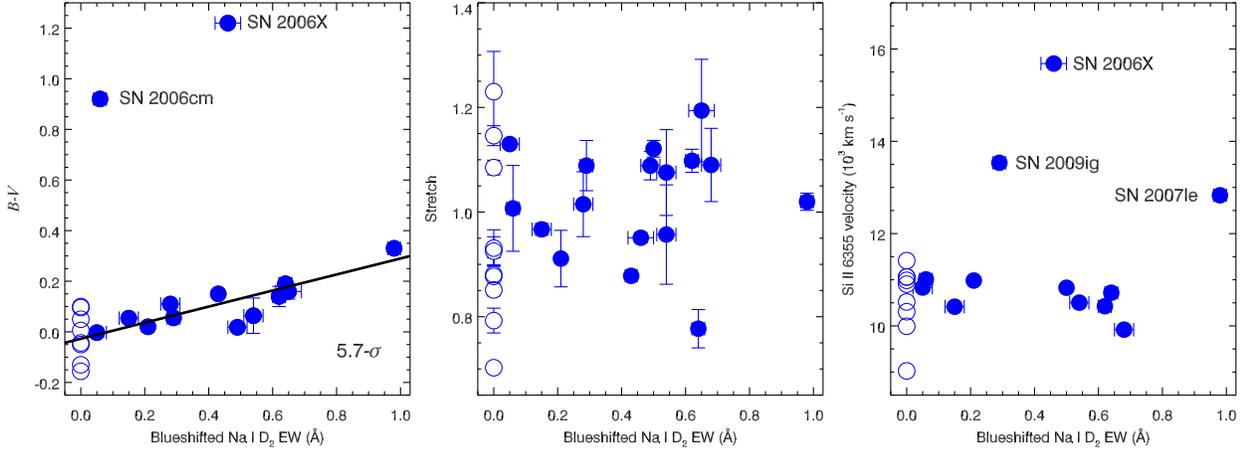}
\caption{\textit{Left:} The $B-V$ colour at maximum as a function of the pEW of the `blueshifted' only narrow \Nai\ D$_2$ features for the SNe Ia in our sample displayed `blueshifted' features. The solid blue circles represent positive detections of   `blueshifted' narrow \Nai\ D$_2$ features, while the open circles represent non-detections. SN\,2006X and SN\,2006cm are extreme outliers to the colour relation and have been excluded from the fit.  The solid black line is the best linear fit to the data, with a significance of 5.7-$\sigma$.   \textit{Middle:} The stretch of the SNe Ia as a function of the pEW of the `blueshifted' only narrow \Nai\ D$_2$ features for the SNe Ia in our sample displaying `blueshifted' features. No correlation is identified. \textit{Right:} The \SiII\ 6355 \AA\ velocity against the  the pEW of the `blueshifted' only narrow \Nai\ D$_2$ features. The three SNe Ia falling in the `HV' group of  \protect \cite{2013arXiv1303.2601W}  are labelled. No correlation is identified.}
\label{naid_ew_sc}
\end{figure*}

 The left panel of Fig.~\ref{naid_ew_sc} shows the $B-V$ colour at maximum as a function of the pEW of their `blueshifted' \Nai\ D$_2$ features. Two SNe Ia in the sample (SN\,2006X, SN\,2006cm) are found to lie nearly 6-$\sigma$ away from the mean $B-V$ colour of the sample and off the identified relation between  \Nai\ D$_2$ pEW and $B-V$ colour at maximum, suggesting a strong contribution from dust in the galaxy to their $B-V$ colour. Excluding these reddened events, we find a strong correlation between the $B-V$ colour at maximum and the `blueshifted' \Nai\ D$_2$ features at the 5.7-$\sigma$ level. 
 
We also investigate the stretch and \SiII\ 6355 \AA\ velocity at maximum as a function of `blueshifted' \Nai\ D$_2$ pEW (also shown in Fig.~\ref{naid_ew_sc}). No statistically significant correlations between either \SiII\ 6355 \AA\ velocity or stretch and `blueshifted' \Nai\ D$_2$ pEW is found.

\section{Discussion}
\label{discussion}

We have presented a new SN Ia spectral sample exploring the connection
between the narrow absorption lines of \Caii\ H\&K and \Nai\ D in SN Ia spectra and the photometric and spectral
properties of SNe Ia. We now discuss the interpretation of these
results in the context of SN Ia progenitor channels, as well as highlight the increasing evidence for distinct families of SNe Ia.

\subsection{Circumstellar material and SN Ia progenitor scenarios}
\label{sec:CSM_disc}

Time-varying blueshifted narrow \Nai\ D absorption features have been identified in some SNe Ia \citep[e.g.,][]{2007Sci...317..924P,2009ApJ...702.1157S,2009ApJ...693..207B,2010AJ....140.2036S}, with the suggestion that these varying profiles are related to outflowing material from their progenitor systems. The velocities of this outflowing material with respect to the defined rest-frame of the SNe are typically $\sim$50--200 \kms, while the distances to the absorbing material have been estimated to be 10$^{16}$--10$^{17}$ cm and having cloud densities of 10$^7$ cm$^{-3}$ \citep{2007Sci...317..924P,2009ApJ...702.1157S}. \citet{2011A&A...530A..63P} showed that the recurrent nova system, RS Ophiuchi (RS Oph), showed very similar time-variable \Nai\ D absorption features during outburst, to those observed in SN\,2006X, suggesting a strong connection between recurrent novae and SNe Ia that show time-variable \Nai\ D absorption features. Three-dimensional modelling of RS Oph has shown that the CSM is expected to be concentrated in the binary orbital plane, suggesting that the probability of detecting CSM is also strongly dependent on viewing angle \citep{2013IAUS..281..195M}.

Regions of very recent star formation ($<$100 Myr) have been found to have outflows with velocities of $\sim$100--200 \kms\ \citep{2013A&A...550A.108V}, similar to those seen in our SN Ia sample. However, SNe Ia are not preferentially found near these regions of recent star formation and therefore, should not be related to these regions of outflowing material -- even so-called `prompt' SNe Ia that trace the host galaxy star formation rate occur on time scales of at least 200 Myr \citep{2009ApJ...707...74R}.

A statistical study of a sample of single-epoch high-resolution spectra of SNe Ia was performed by \citet{2011Sci...333..856S}, where an excess of SNe Ia displaying blueshifted \Nai\ D absorption features compared to non-blueshifted \Nai\ absorption features in spiral galaxies was found. This was interpreted as evidence favouring
the SD progenitor channel for some SNe Ia in spiral galaxies.  

In our sample, we find an excess of SNe Ia displaying blueshifted \Nai\ D absorption features compared to those that show non-blueshifted absorption features, with an estimate of $\sim$20 per cent of SNe Ia showing additional blueshifted \Nai\ D absorption features. This value is most likely a lower limit since the CSM distribution is expected to be asymmetric, resulting in cases where CSM is present but not seen due to the viewing angle. We also show that the presence of \Nai\ D absorption features is strongly dependent on host galaxy properties with SNe Ia displaying blueshifted features predominantly found in later-type galaxies. No SN Ia with blueshifted \Nai\ D features is found in an early-type (E/S0) galaxy. Although \Nai\ D and \Caii\ H\&K absorption lines arising from within the host galaxies are expected to be stronger in late-type galaxies, there is no simple `host galaxy' explanation for the observed excess of `blueshifted' narrow absorption components.

We also find that the strength of \Nai\ D absorption features is correlated with the presence or absence of blueshifted \Nai\ D absorption features -- SNe Ia with blueshifted \Nai\ D absorption features have stronger \Nai\ D lines (measured through the pEW of the stronger of the two \Nai\ D lines,  \Nai\ D$_2$)  than SNe Ia that do not have blueshifted \Nai\ D features.  This increased \Nai\ D absorption depth in SNe Ia showing `blueshifted' material, coupled with the identified excess of SNe Ia with `blueshifted' SNe Ia, is strongly suggestive of an  additional contribution to the \Nai\ D absorption from the SN progenitor system.

SNe Ia displaying no \Nai\ D absorption features have been found to have narrower light curves than those that show `blueshifted' \Nai\ D absorption features \citep{2012ApJ...752..101F}. Our data suggests that SNe Ia showing \Nai\ D absorption features (both `blueshifted' and `non-blueshifted) have broader light curves than those that do not show \Nai\ D absorption features. However, interestingly we find no correlation between the pEW strength of `blueshifted' \Nai\ D$_2$ features and the light curve width, suggesting that once we look within the sample of SNe Ia with `blueshifted' \Nai\ D$_2$ features, there is no additional relation between \Nai\ D strength and light curve width. These results appears to be linked to the host galaxy distribution of the sample -- SNe Ia with  \Nai\ D absorption features are predominantly found in late-type host galaxies, where it is well known that the SNe Ia display broader light curves \citep{1995AJ....109....1H,1996AJ....112.2391H,1999AJ....117..707R,2000AJ....120.1479H}.  However, it is still difficult to identify the driving force behind these correlations since the reason why less luminous SNe Ia preferentially occur in early-type galaxies compared to late-type galaxies is not well understood.

SNe Ia with blueshifted \Nai\ D absorption features have previously been suggested to have redder $B-V$ colours at maximum and higher  \SiII\ 6355 \AA\ velocities compared to the SN Ia population as a whole \citep{2012ApJ...752..101F}. The three SNe Ia in our combined sample with the highest velocities ($>12000$ \kms), and falling in the `high-velocity' class of \cite{2009ApJ...699L.139W}, do display blueshifted absorption features, and the two reddest SNe Ia in the sample display blueshifted \Nai\ D features.  However, for our sample, we find no statistically significant difference between the velocity and colour of the SN Ia sample when split into those with blueshifted \Nai\ D absorption profiles and those without. We also find that these differences are primarily driven by SN\,2006X, a very red outlier with high \SiII\ velocities, which has blueshifted \Nai\ D absorption features. The highly interacting SN Ia, PTF11kx, which displayed very strong H in its spectra, as well as time-varying \Nai\ D absorption, had a relatively low \SiII\ 6355 \AA\ velocity at maximum of $\sim$11000 \kms, which does not suggest a continuing trend of higher velocities towards more highly interacting SNe Ia. We also find no correlation between \SiII\ 6355 \AA\ velocity at maximum and the strength (pEW) of `blueshifted' \Nai\ D absorption features.

For completeness, we note that none of the SNe Ia with `HV' \SiII\ features are from the new XShooter sample nor does the XShooter sample contain any SN Ia with a red \textit{B-V} colours that would exceed the colour cutoff used in SN Ia cosmological studies of 0.25, while the S11 sample contains four. However, we do not find a statistically significant difference between the XShooter and S11 colour or \SiII\ velocity distributions.

When we study the connection between the pEW of `blueshifted \Nai\ D$_2$ features (pEW of features blueshifted with respected to the zero velocity) and the SN $B-V$ colour at maximum light, we find a strong correlation  (5.7-$\sigma$). An obvious explanation for SNe Ia with stronger `blueshifted' \Nai\ D pEW  having redder colours is that dust in the CSM makes an additional contribution to the extinction towards the SN resulting in a redder $B-V$ colour. The effect of circumstellar dust on the B-V colour and the effective extinction law has been previously modelled \citep{2008ApJ...686L.103G,2011ApJ...735...20A}. It has been found that dust in the CSM could cause $B-V$ colour variations are of the order of 0.05--0.1 mag for dust radii of 10$^{16}$--10$^{19}$ cm.

\cite{2013ApJ...772...19F} have recently shown that SNe Ia that are faster Lira law $B-V$ decliners  (35--80 d after maximum) have higher pEW values of unresolved \Nai\ D absorption features (measured from lower resolution spectra) and redder colours at maximum than slower Lira law $B-V$ decliners. They attribute this difference as evidence of CSM in the `fast decliner' group. A direct comparison with their results cannot be made since their study uses lower resolution spectra, and therefore, information on the relative velocity shifts and strength of the `blueshifted' \Nai\ D features is not available.

\subsection{CSM from double degenerate channels}
\label{DD}

Some recent work has attempted to explain the observations of outflowing CSM material using a DD origin. \citet{2013arXiv1302.2916S} showed that the interaction between material ejected from a He-CO WD binary system and the ISM could produce outflowing neutral Na.  In agreement with our results, they also found that the
lower ISM densities in elliptical galaxies would inhibit detection of blueshifted absorption features in these galaxies.  \citet{2013arXiv1304.4957R} showed that tidal tails from double degenerate WD mergers interacting with the ISM may also produce outflowing material that could result in observed blueshifted absorption features. Additional analysis and simulations are necessary to quantify the absorbing material that would be present and to explore the exact conditions of the CSM and ISM that are needed to produce the observed features and correlations with SN properties.

Models of the violent merger of a WD with the core of a giant star during the common envelope phase can also explain the presence of CSM \citep{2013MNRAS.431.1541S}. Population synthesis modelling has suggested that some SNe Ia produced from violent CO+CO WD mergers may show signs of CSM if the SN Ia explodes soon after the common envelope phase  ($<1000$ yr), although the rate of these events is expected to be low \citep[$<$4 per cent,][]{2013MNRAS.429.1425R}. Models of canonical CO+CO WD mergers require delay times of 10$^5$ yr between the initial dynamical merger and the explosion, effectively ruling out this CSM production mechanism in non-violet WD mergers \cite[][]{2007MNRAS.380..933Y}.
Therefore, it seems unlikely  that this channel can fully explain the rates of blueshifted
\Nai\ D absorption features that have been observed. 

\subsection{Evidence for two (or more) families of SNe Ia?}
\label{sec:evid-two-famil}

\begin{table}
 \caption{Observational evidence for two families of SNe Ia.}
 \label{thetable}
 \begin{tabular}{@{}lccccccccccccccccccccccccccccccc}
  \hline
  \hline
Family 1& Family 2&Ref.\\
\hline
More luminous & Less luminous&1 \\
Broad light curve & Narrow light curve&1\\
Low \SiII\ 4130 pEW& High \SiII\ 4130 pEW &2\\
Stronger high-velocity features & Weaker high-velocity features &3\\
Late-type host& Early-type host&4\\
Low M$_{stellar}$ host & High M$_{stellar}$ host&5\\
High sSFR host & Low sSFR host&6 \\
Short delay-time & Long delay-time&7\\
Blueshifted \Nai\ D absorption & No \Nai\ D absorption&this paper\\
\hline
 \end{tabular}
\begin{flushleft}
 $^1$\cite{1993ApJ...413L.105P}\\
 $^2$\cite{2008A&A...477..717B,2008A&A...492..535A,2011MNRAS.410.1262W,2011A&A...526A..81B,2011A&A...529L...4C,2012AJ....143..126B, 2012MNRAS.425.1889S}\\
 $^3$\cite{2012MNRAS.426.2359M,2013arXiv1307.0563C}\\
 $^4$\cite{1995AJ....109....1H,1996AJ....112.2391H,1999AJ....117..707R,2000AJ....120.1479H}\\
$^5$\cite{2010MNRAS.406..782S}\\
$^6$\cite{2005ApJ...634..210G,2006ApJ...648..868S}\\
$^7$\cite{2005A&A...433..807M,2006MNRAS.370..773M,2005ApJ...629L..85S,2006ApJ...648..868S,2012ApJ...755...61S,2013ApJ...770..108C}\\
\end{flushleft}
\end{table}

The observed connection between the presence (and strength) of narrow blueshifted \Nai\ D absorption features in SN spectra and observed SN properties has a number
of implications, independent of the exact progenitor configuration. The clear excess of SNe Ia in our sample displaying blueshifted \Nai\ D absorption compared to those with non-blueshifted \Nai\ D absorption suggests an additional contribution from CSM to the absorption profiles. SNe Ia with wider light curves have long been known to be intrinsically
more luminous \citep{1993ApJ...413L.105P}, but many other SN
properties are also connected to light curve width. 

In Table \ref{thetable}, we show the observed SN Ia properties that are found to be related to SN Ia luminosity. As well as broader light curves ($s>1.0$)\footnote{The position of the split between low and high stretch is made based on the mean stretch value of the SNLS SN Ia sample (s=1.0) from \cite{2010A&A...523A...7G}.}, more luminous SNe Ia also have lower \SiII\ 4130 \AA\ pEW than less luminous SNe Ia \citep{2008A&A...477..717B,2008A&A...492..535A,2011MNRAS.410.1262W,2011A&A...526A..81B,2011A&A...529L...4C,2012AJ....143..126B}.  

Trends of some SN spectral feature velocities increasing with increasing light curve width have previously been identified \citep[e.g.,][]{1994AJ....108.2233W,1995ApJ...447L..73F, 2007Sci...315..825M,2012MNRAS.426.2359M} -- higher \Caii\ H\&K and \Caii\ NIR velocities at maximum are associated with SNe Ia with broader light curve widths. One possible cause for these correlations is a stronger contribution from `high-velocity' material in more luminous SNe Ia \citep{2012MNRAS.426.2359M}. \cite{2013arXiv1307.0563C} have identified an connection between the strength of `high-velocity' features in SNe Ia spectra and light curve width, with more luminous SNe Ia displaying stronger `high-velocity' components in the \Caii\ NIR feature. One suggested mechanism for `high-velocity' features is a density enhancement at high velocity, either from circumstellar material (i.e. a `detached shell' of material) or intrinsic to the SN \citep{2005ApJ...623L..37M,2012AJ....143..126B}. 

The photometric properties of SNe Ia are also closely linked to
their host galaxy properties. SNe Ia in morphologically elliptical
systems are intrinsically fainter and lower stretch than SNe Ia in
spiral or late-type galaxies
\citep{1995AJ....109....1H,1996AJ....112.2391H,1999AJ....117..707R,2000AJ....120.1479H}.
These correlations also apply when examining physical variables, which
define the stellar populations of the SN hosts instead of morphology,
with more luminous, higher stretch SNe in less massive galaxies
\citep{2010MNRAS.406..782S} and more strongly star-forming galaxies
\citep{2005ApJ...634..210G,2006ApJ...648..868S} than less luminous (lower stretch) SNe Ia. Thus high stretch SNe Ia
seem to favour younger stellar environments and thus likely younger
progenitor systems.  X-rays signatures of accreting WD systems in early-type galaxies have also been found to be rare with \cite{2010Natur.463..924G} suggesting that no more than 5 per cent of SNe Ia in early-type galaxies can come from the SD channel \citep[although see ][]{2011RAA....11..965M,2013IAUS..281..145K}.

Direct measurements of the SN Ia host galaxy ages, although very
challenging, also support a relation between stretch and galaxy age
\citep[][Pan et al. in preparation]{2012arXiv1211.1386J}. This is supported by studies of the SN Ia
delay-time distribution (DTD); when the DTD is constructed for low and
high stretch events, again higher stretch SNe favour a younger DTD
\citep{2010AJ....140..804B,2012MNRAS.426.3282M}. SN Ia rates have also been found to depend on both host galaxy star formation rate and host stellar mass, leading to the idea of a two-component model for SNe Ia, with both `prompt' ($<$500 Myr) and `delayed' ($>$500 Myr) channels acting to produce SNe Ia \citep{2005A&A...433..807M,2006MNRAS.370..773M,2005ApJ...629L..85S,2006ApJ...648..868S,2012ApJ...755...61S,2013ApJ...770..108C}.

\cite{2013arXiv1303.2601W} have suggested a connection between the \SiII\ 6355 \AA\ velocity at maximum and the location of SNe Ia in their host galaxies -- more centrally located SNe Ia display, on average, higher \SiII\ 6355 \AA\ velocities, suggested to be linked to a younger stellar population. However, they find no connection between \SiII\ 6355 \AA\ velocity and light curve width, which may be expected given the previously identified links between younger stellar populations and higher SN luminosity. 

SN Ia remnants also show diversity in their properties with Galactic SN remnants such as Tycho showing no signs of interaction with circumstellar material \citep{2006ApJ...645.1373B}, while searches for a non-degenerate companion star for Tycho have not identified a companion star consistent with its expected properties \citep{2012arXiv1210.2713K}. However, Kepler's SN remnant, which has been shown to be an over-luminous SN Ia \citep{2012ApJ...756....6P} displays signatures of interaction between the SN ejecta and circumstellar material \citep[e.g.,][]{1987ApJ...319..885B,1982A&A...112..215D,2013ApJ...764...63B}, as well as strong evidence for super-solar metallicity of the progenitor \citep{2013ApJ...767L..10P} and warm dust originating from swept-up circumstellar material \citep{2012MNRAS.420.3557G}. These properties are consistent with the explosion of a SN Ia through a relatively prompt progenitor channel, while Tycho is more consistent with an older progenitor.

The connection between SN luminosity and SN observables presented in Table \ref{thetable} is suggestive of a split in SNe Ia properties based on their luminosity, which correlates with many observed SN Ia properties.  Within `Family 1', we have further identified a trends between SN $B-V$ colour at maximum and pEW of the `blueshifted' \Nai\ D absorption features, suggestive of an increasing contribution from CSM leading to an increasing contribution from dust. 

However, it is still unclear what the driving force behind these correlations is and whether they are directly linked to independent progenitor channels. Any viable  progenitor model must be able to explain this observed diversity in SN Ia properties. Our estimate is that at least 20 per cent of SNe Ia have an additional contribution from CSM as measured from the excess of `blueshifted' \Nai\ D absorption features, as well as their stronger \Nai\ D pEW than the rest of the SN Ia sample. We interpret this percentage as a lower limit since any asymmetry in the CSM \cite[i.e. a concentration in the binary orbital plane as shown in RS Oph; ][]{2013IAUS..281..195M} will result in line-of-sight effects, where outflowing CSM is present but not observed.

\subsubsection{Type Ia-CSM SNe}

For `normal' SNe Ia the detection of H in their spectra is a
key diagnostic of the SD progenitor channel, since the origin of this
H is most likely from a non-degenerate companion star. This detection
of hydrogen lines had, until recently, only been confirmed for a
handful of events: SN\,2002ic \citep{2003Natur.424..651H,2004ApJ...605L..37D,2004MNRAS.354L..13K,2004ApJ...604L..53W,2004ApJ...616..339W}, SN\,2005gj
\citep{2006ApJ...650..510A,2007arXiv0706.4088P}, SN\,2008J \citep{2012A&A...545L...7T}, and
PTF11kx \citep{2012Sci...337..942D}. These objects
are interacting with their CSM to an even stronger degree than the SNe
studied here, with strong narrow hydrogen emission, and have recently
been dubbed `SNe Ia-CSM' \citep{2013arXiv1304.0763S}. These authors
conducted a detailed search in the spectra of SNe classified as IIn
from PTF and the literature, for diluted signatures of SNe Ia, and
found a total of 16 Ia-CSM objects. 

The most common SN Ia template spectrum for the SNe Ia-CSM was suggested by  \cite{2013arXiv1304.0763S} to be, the brighter than average, 1991T/1999aa-like spectrum. However, it is only with model spectral templates that this has been shown not to be a luminosity bias \citep{2013arXiv1306.1549L}. While SN 1991T itself did not show the narrow emission lines typical of SNe Ia-CSM and IIn, it did display blue-shifted \Nai\ D absorption features \citep{2007Sci...317..924P}, suggesting an association between SN\,1991T and the more luminous than average SNe Ia that show strong signatures of CSM interaction.

In some respects, events in this SN Ia-CSM class may represent a more
extreme version of the objects observed here, and are also
preferentially located in late-type spirals or dwarf irregulars,
star-forming galaxies similar to the hosts of our blueshifted \Nai\ D absorption feature
objects.

\subsubsection{The search for H in late time spectra}
\label{sec:search-h-late}

Even for SN Ia events without a dense enough CSM to produce hydrogen in the
photospheric spectra, hydrogen-rich material may still be stripped
from the companion star by the SN ejecta. This hydrogen is then
predicted to be present at low velocities ($\sim$1000 \kms) and can
only be detected after the outer (higher velocity) layers have become
optically thin \citep{2000ApJS..128..615M, 2007PASJ...59..835M}. A
search for this hydrogen has been conducted in five `normal' SNe Ia at the necessary late times
(SN\,2001el, SN\,2005am, SN\,2005cf, SN\,2011fe, SN 1998bu), but none has yet been detected,
with a mass limit of the hydrogen of $<$0.03 \msun ($<$0.001 \msun\ for nearby SN\,2011fe), disfavouring a
single-degenerate progenitor channel
\citep[][]{2005A&A...443..649M,2007ApJ...670.1275L, 2013ApJ...762L...5S,2013arXiv1307.4099L}. However, some models predict that the companion star will have lost its envelope by the time of explosion, providing an alternative explanation for the absence of H features in SNe Ia \citep{2011ApJ...730L..34J,2011ApJ...738L...1D}.

However, these SNe Ia have stretches of 0.99$\pm$0.01 (SN\,2001el),
0.96$\pm$0.01 (SN\,2005cf), 0.70$\pm$0.03 (SN\,2005am), 0.96$\sim$0.02 (SN\,1998bu), and 0.98$\pm$0.01 (SN\,2011fe), placing
them all in the lower-stretch family ($s<1.01$). If the hypothesis that this lower luminosity family comes from an older stellar population such as through the DD progenitor channel, we do not expect to see hydrogen in their late-time spectra. A search
for hydrogen in the spectra of higher stretch SNe Ia would therefore
be an interesting study.

\subsubsection{Individual case studies}
\label{sec:indiv-case-stud}

Finally, we discuss two well-studied SNe Ia from our sample to determine which
column of Tab.~\ref{thetable} they most likely fit into.  However, we
stress that since some SNe Ia show blueshifted \Nai\ D absorption features associated with host galaxy ISM and not CSM, it is not necessary that
individual objects will fall completely into either class.

SN\,2011fe exploded in M101, a high stellar mass galaxy
\citep{2011Natur.480..344N}.  Its light curve stretch of 0.98 and lower
spectral velocities place it in Family 2.  Despite extensive
monitoring, no CSM was detected in high-resolution spectra.
Pre-explosion imaging has ruled out more massive companion progenitor
stars, favouring a low-mass or degenerate companion star
\citep{2011Natur.480..348L}. As discussed in Section \ref{sec:search-h-late}, no H was detected in its late-time spectra with an upper mass limit of $<$0.001 \msun.

SN\,2012cg was a nearby ($\sim$15 Mpc) SN Ia discovered just 1.5$\pm$0.2 d
after explosion. It had a stretch of 1.098$\pm$0.022, occurred in a barred Sa galaxy (typically low stellar mass, high sSFR) and displayed blueshifted narrow absorption features of \Caii\ H\&K and \Nai\ D. These characteristics place it in Family 1 of Tab.~\ref{thetable}. 

\section{Conclusions}
\label{sec:conclusions}

In this paper, we have presented a sample of 17 low-redshift SNe Ia observed with the XShooter intermediate resolution spectrograph on the VLT. We conducted a search for narrow \Nai\ D absorption profiles in these
spectra and, where present, have measured its blueshift (or non-blueshift)
 relative to the systemic velocity of the SN in its host
galaxy.  We combined these new data with events from the literature to
form a single sample of 32 SNe Ia with intermediate-high resolution
spectra and light curve data. We also measured the strength of the narrow \Nai\ D absorption features through pEW measurements and investigated the connection to SN observables. Our main conclusions are:

\begin{enumerate}
\item Combining our new data with the S11 sample, we find an excess of SNe Ia with blueshifted \Nai\ D absorption features over those with no-blueshifted \Nai\ D, with $\sim$20 per cent of SNe Ia having an additional blueshifted \Nai\ D absorption feature.

\item SNe Ia with \Nai\ D absorption features in their spectra
  have, on average, broader light curves (or higher stretches) and are
  more luminous events than SNe Ia without \Nai\ D absorption features.
    
\item SNe Ia with blueshifted \Nai\ D absorption features are most likely to be found in late-type galaxies containing younger stellar populations. No SNe Ia in our sample with blueshifted \Nai\ D were  found in an E/S0 galaxy.
  
\item SNe Ia with blueshifted \Nai\ D absorption features show stronger \Nai\ D pEWs than those without blueshifted features, suggestive of an additional contribution to the \Nai\ D absorption from CSM.

\item Within the sample of SNe Ia with blueshifted \Nai\ D absorption, we find that the strength of the `blueshifted' \Nai\ D absorption features correlates with SN $B-V$ colour at maximum, strongly suggesting this material is associated with the progenitor system.
  
\item We find no statistically significant preference for SNe Ia with blueshifted \Nai\ D features to have higher \SiII\ 6355 \AA\ velocities than SNe Ia without blueshifted \Nai\ D features.
  \end{enumerate}

The simplest explanation for the presence of additional blueshifted \Nai\ D absorption features in SN Ia
spectra is that it arises due to CSM from the progenitor system of the SN. This suggests a progenitor channel where one would expect outflowing shell-like structures  - the most obvious being the SD scenario. A SD origin for the CSM is supported by clear observational evidence with recurrent nova systems being observed to show time-varying \Nai\ D features very similar to those in some SNe Ia. However, some recent DD models may now also
produce similar narrow absorption features, but not currently at the rate necessary to explain our
results.

Table \ref{thetable} summarises the observational evidence for two
distinct families of `normal' SNe Ia with different light curve,
spectral, and host galaxy properties. The rates of the different
channels are consistent with one population with high luminosity, short delay-times and
evidence for outflowing material, with the other population displaying no \Nai\ D absorption features, low luminosity and long delay-times indicative of an older population.  Whether these `families' correspond to separate progenitor channels (SD and DD) or can be explained within the framework of one channel (i.e. different types of companion stars in the SD channel) is still very much a topic under debate.

\section{Acknowledgements}
MS acknowledges support from the Royal Society.  Research leading to these results has received funding from the European Research Council (ERC) under the European UnionÕs Seventh Framework Programme (FP7/2007- 2013)/ERC Grant agreement no [291222] (PI : S. J. Smartt). A.G. is supported by the EU/FP7 via an ERC grant, and by Minerva/ARCHES and Kimmel awards.  SH is supported by a Minerva/ARCHES award. GL is supported by the Swedish Research Council through grant No. 623-2011-7117.

Based on observations collected at the European Organisation for Astronomical Research in the Southern Hemisphere, Chile as part of PESSTO, (the Public ESO Spectroscopic Survey for Transient Objects Survey) ESO program ID 188.D-3003, as well as observations made with ESO Telescopes at the La Silla Paranal Observatory under programme ID 090.D-0828(A) and 089.D-0647(A). Based on observations (GS-2012B-Q-86) obtained at the Gemini Observatory, which is operated by the Association of Universities for Research in Astronomy, Inc., under a cooperative agreement with the NSF on behalf of the Gemini partnership: the National Science Foundation (United States), the National Research Council (Canada), CONICYT (Chile), the Australian Research
Council (Australia), Minist\'erio da Ci\^encia, Tecnologia e Inova\c{c}\~ao (Brazil) and Ministerio de Ciencia, Tecnolog\'ia e Innovaci\'on Productiva (Argentina). Based in part on data from the 1.3m telescope operated by the SMARTS consortium. 

The Liverpool Telescope is operated on
the island of La Palma by Liverpool John Moores University in the
Spanish Observatorio del Roque de los Muchachos of the Instituto de
Astrofisica de Canarias with financial support from the UK Science and
Technology Facilities Council.  Observations were obtained with the
Samuel Oschin Telescope at the Palomar Observatory as part of the
Palomar Transient factory project, a scientific collaboration between
the California Institute of Technology, Columbia Unversity, La Cumbres
Observatory, the Lawrence Berkeley National Laboratory, the National
Energy Research Scientific Computing Center, the University of Oxford,
and the Weizmann Institute of Science. The William Herschel Telescope
is operated on the island of La Palma by the Isaac Newton Group in the
Spanish Observatorio del Roque de los Muchachos of the Instituto de
Astrofísica de Canarias.  This paper uses observations obtained with facilities of
the Las Cumbres Observatory Global Telescope.  This research has made use of the NASA/IPAC Extragalactic Database (NED) which is operated by the Jet Propulsion Laboratory, California Institute of Technology, under contract with the National Aeronautics and Space Administration. 

\bibliographystyle{mn2e}
\bibliography{astro}

\appendix
\label{appa}
\section{S11 sample values}

\begin{table*}
  \caption{Information for S11 sample including light curve and spectral measurements. \SiII\ 6355 \AA\ velocities are measured SN line velocities in the phase range -2 to +5 d with respect to \textit{B}-band maximum. The references for the photometric and spectral properties are given in Sec.~\ref{sec:literature-data}.}
 \label{tab:appda}
\begin{tabular}{@{}lccccccccccccccccccccccccccccc}
  \hline
  \hline
 SN&$z^{helio}$&Stretch&\textit{B-V} &\Nai\ D$_2$ &`Blueshifted'&Galaxy& \SiII\ 6355 \AA\ &Phase$^i$ \\
&&&at max. &pEW (\AA)&\Nai\ D$_2$ pEW (\AA)$^{h}$& type&vel. (10$^3$ \kms) &(d)  \\
\hline
\hline
 &&&\textbf{Blueshifted \Nai\ D}\\
 \hline
 SN\,2006X&0.005486$\pm$0.000006$^a$&0.950$\pm$0.010&1.22$\pm$0.01&1.16$\pm$0.03&0.46$\pm$0.04&Sbc&15.68$\pm$0.10&0&\\
SN\,2006cm$^*$&0.016434$\pm$0.000007$^b$&1.007$\pm$0.082&0.92$\pm$0.02&1.10$\pm$0.01&0.06$\pm$0.02&Sb&11.01$\pm$0.13&0\\
SN\,2007af$^*$&0.00547$^c$&0.967$\pm$0.011&0.05$\pm$0.01&0.20$\pm$0.02&0.15$\pm$0.03&Scd&10.42$\pm$0.11&0\\
SN\,2007le&0.00713$^d$&1.020$\pm$0.016&0.33$\pm$0.01&0.98$\pm$0.02&0.98$\pm$0.02&Sc&12.83$\pm$0.13$^*$&0$^j$\\
SN\,2008C&0.016925$\pm$0.000007$^b$&0.777$\pm$0.037&0.19$\pm$0.02&0.64$\pm$0.02&0.64$\pm$0.02&S0/a&10.72$\pm$0.15&+2\\
SN\,2008ec&0.016317$\pm$0.000007$^e$&0.878$\pm$0.011&0.15$\pm$0.01&0.43$\pm$0.01&0.43$\pm$0.01&Sa&--&--\\
SN\,2009ig&0.008770$\pm$0.000021$^e$ &1.089$\pm$0.048&0.06$\pm$0.01&0.29$\pm$0.02&0.29$\pm$0.02&Sa& 13.53$\pm$0.13&0\\
SN\,2009le$^*$&0.017458$\pm$0.000007$^f$&1.076$\pm$0.082&0.06$\pm$0.07&1.16$\pm$0.02&0.54$\pm$0.03&Sbc&--&--\\
SN\,2009ds$^*$&0.019032$\pm$0.000007$^g$ &1.089$\pm$0.027&0.02$\pm$0.02&0.66$\pm$0.01&0.49$\pm$0.03&Sc&--&--\\
SN\,2010A$^*$&0.020698$\pm$0.000033$^e$ &1.015$\pm$0.062&0.11$\pm$0.02&0.47$\pm$0.01&0.28$\pm$0.03&Sab&--&--\\
\hline
 &&&\textbf{No blueshifted \Nai\ D}\\
\hline
SN\,2007kk&0.041045$\pm$0.000103$^e$&1.098$\pm$0.041&0.00$\pm$0.02&0.47$\pm$0.01&--&Sbc&--&--\\
SN\,2008fp&0.005664$\pm$0.000067$^e$&1.067$\pm$0.020&0.58$\pm$0.02&1.20$\pm$0.01&--&S0 pec&10.83$\pm$0.10&0\\
\hline
&&&\textbf{No \Nai\ D}\\
\hline
SN\,2007on&0.006494$\pm$0.000013$^e$&0.702$\pm$0.007&0.10$\pm$0.01&--&--&E&11.06$\pm$0.12&$-$1\\
SNF20080514-002&0.022095$\pm$0.000090$^e$&0.793$\pm0.024$&-0.16$\pm$0.02&--&--&S0&10.31$\pm$0.16&+3\\
SN\,2008hv&0.012549$\pm$0.000067$^e$&0.851$\pm$0.011&0.01$\pm$0.02&--&--&S0&10.90$\pm$0.12&$-1$\\
SN\,2008ia&0.021942$\pm$0.000097$^e$&0.880$\pm$0.032&0.05$\pm$0.03&--&--&E1&11.06$\pm$0.18&+3\\
\hline
 \hline
\end{tabular}
 \begin{flushleft}
  $^a$Redshift calculated from of CN lines in high-resolution UVES spectrum from \citet{2007Sci...317..924P}. \\
   $^b$Redshift from host galaxy features in SN spectrum. \\
  $^c$Redshift from H$\alpha$ emission from \citet{2007ApJ...671L..25S}. \\
   $^d$Redshift from H$\alpha$ emission from \citet{2009ApJ...702.1157S}. \\
   $^e$Redshift from recessional velocity obtained from NED or SDSS Data Release 9.\\
   $^f$Redshift from H$\alpha$ emission from high-resolution spectrum (Josh Simon, priv. communication).\\
   $^g$Redshift from H$\alpha$ emission from high-resolution spectrum (Assaf Sternberg, priv. communication).\\
   $^h$`Blueshifted' \Nai\ D pEW refers to the integrated pEW of any \Nai\ D absorption features that are blueshifted with respect to the defined zero velocity position.\\
   $^i$Phase with respect to \textit{B}-band maximum, as measured using SiFTO\\
   $^j$Average \SiII\ 6355 \AA\ velocity of -6 d and +6 d spectra. \\
   $^*$SN 2006cm, SN 2007af, SN 2009ds, SN 2009le and SN 2010A are removed from our calculation of the ratio of `blueshifted' to `redshifted' absorption features since they display both `blueshifted' and `non-blueshifted' \Nai\ D absorption components.
   
\end{flushleft}
\end{table*}

\end{document}